\documentclass[pra,twocolumn,showkeys,showpacs,superscriptaddress,floatfix,longbibliography]{revtex4-2}

\usepackage[T1]{fontenc}
\usepackage[latin9]{inputenc}
\usepackage{epsfig,psfrag}
\usepackage{latexsym}
\usepackage{graphicx}
\usepackage{dcolumn}
\usepackage{amssymb}
\usepackage{amsmath}
\usepackage{bm}
\usepackage{mathrsfs}
\usepackage{bigints}
\usepackage{upgreek}
\usepackage{color}
\usepackage{xspace}
\usepackage[squaren]{SIunits}
\usepackage{epstopdf}
\usepackage{float}
\usepackage[normalem]{ulem}
\usepackage{hyperref}

\definecolor{mygreen}{RGB}{20,148,20}

\usepackage{filemod}

\newcommand{\fig}[1]{Fig.~\ref{fig:#1}}

\begin{document}

\title{Bi-color atomic beam slower and magnetic field compensation for ultracold gases}

\author{Jianing Li}
\email{LIJI0042@e.ntu.edu.sg}
\affiliation{Nanyang Quantum Hub, School of Physical and Mathematical Sciences, Nanyang Technological University, 21 Nanyang Link, Singapore 637371, Singapore}
\affiliation{MajuLab, International Joint Research Unit IRL 3654, CNRS, Universit\'e C\^ote d'Azur, Sorbonne Universit\'e, National University of Singapore, Nanyang Technological University, Singapore}
\author{Kelvin Lim}
\affiliation{Nanyang Quantum Hub, School of Physical and Mathematical Sciences, Nanyang Technological University, 21 Nanyang Link, Singapore 637371, Singapore}
\affiliation{MajuLab, International Joint Research Unit IRL 3654, CNRS, Universit\'e C\^ote d'Azur, Sorbonne Universit\'e, National University of Singapore, Nanyang Technological University, Singapore}
\author{Swarup Das}
\affiliation{Nanyang Quantum Hub, School of Physical and Mathematical Sciences, Nanyang Technological University, 21 Nanyang Link, Singapore 637371, Singapore}
\affiliation{MajuLab, International Joint Research Unit IRL 3654, CNRS, Universit\'e C\^ote d'Azur, Sorbonne Universit\'e, National University of Singapore, Nanyang
Technological University, Singapore}
\author{Thomas Zanon-Willette}
\affiliation{MajuLab, International Joint Research Unit IRL 3654, CNRS, Universit\'e C\^ote d'Azur, Sorbonne Universit\'e, National University of Singapore, Nanyang
Technological University, Singapore}
\affiliation{Centre for Quantum Technologies, National University of Singapore, 117543 Singapore, Singapore}
\affiliation{{Sorbonne Universi\'e, Observatoire de Paris, Universit\'e PSL, CNRS, LERMA, F-75005, Paris, France}}
\author{Chen-Hao Feng}
\affiliation{LP2N, Laboratoire Photonique, Num\'erique et Nanosciences, Universit\'e Bordeaux-IOGS-CNRS:UMR 5298, F-33400 Talence, France}
\author{Paul Robert}
\affiliation{LP2N, Laboratoire Photonique, Num\'erique et Nanosciences, Universit\'e Bordeaux-IOGS-CNRS:UMR 5298, F-33400 Talence, France}
\author{Andrea Bertoldi}
\affiliation{LP2N, Laboratoire Photonique, Num\'erique et Nanosciences, Universit\'e Bordeaux-IOGS-CNRS:UMR 5298, F-33400 Talence, France}
\author{Philippe Bouyer}
\affiliation{LP2N, Laboratoire Photonique, Num\'erique et Nanosciences, Universit\'e Bordeaux-IOGS-CNRS:UMR 5298, F-33400 Talence, France}
\author{Chang Chi Kwong}
\affiliation{Nanyang Quantum Hub, School of Physical and Mathematical Sciences, Nanyang Technological University, 21 Nanyang Link, Singapore 637371, Singapore}
\affiliation{MajuLab, International Joint Research Unit IRL 3654, CNRS, Universit\'e C\^ote d'Azur, Sorbonne Universit\'e, National University of Singapore, Nanyang
Technological University, Singapore}
\author{Shau-Yu Lan}
\affiliation{Nanyang Quantum Hub, School of Physical and Mathematical Sciences, Nanyang Technological University, 21 Nanyang Link, Singapore 637371, Singapore}
\affiliation{MajuLab, International Joint Research Unit IRL 3654, CNRS, Universit\'e C\^ote d'Azur, Sorbonne Universit\'e, National University of Singapore, Nanyang
Technological University, Singapore}
\author{David Wilkowski}
\affiliation{Nanyang Quantum Hub, School of Physical and Mathematical Sciences, Nanyang Technological University, 21 Nanyang Link, Singapore 637371, Singapore}
\affiliation{MajuLab, International Joint Research Unit IRL 3654, CNRS, Universit\'e C\^ote d'Azur, Sorbonne Universit\'e, National University of Singapore, Nanyang
Technological University, Singapore}
\affiliation{Centre for Quantum Technologies, National University of Singapore, 117543 Singapore, Singapore}

\begin{abstract}
Transversely loaded bidimensional-magneto-optical-traps (2D-MOTs) have been recently developed as high flux sources for cold strontium atoms to realize a new generation of compact experimental setups. Here, we discuss on the implementation of a cross-polarized bi-color slower for a strontium atomic beam, improving the 2D-MOT loading, and increasing the number of atoms up to $\sim 10^9$ atoms in the 461 nm MOT. Our slowing scheme addresses simultaneously two excited Zeeman substates of the $^{88}$Sr $^1$S$_0\rightarrow\,^1$P$_1$ transition at 461~nm. We also realized a three-axis active feedback control of the magnetic field down to the microgauss regime. Such a compensation is performed thanks to a network of eight magnetic field probes arranged in a cuboid configuration around the atomic cold sample, and a pair of coils in a quasi-Helmholtz configuration along each of three Cartesian directions. Our active feedback is capable of efficiently suppressing most of the magnetically-induced position fluctuations of the 689~nm intercombination-line MOT.
\end{abstract}



\maketitle

\section{Introduction}

{Since the first demonstration of laser cooling and trapping~\cite{Metcalf_lasercooling,Pritchard_MOT}, cold and ultracold trapped atoms have played a vital role in realizing many advanced atomic physics experiments through the coherent controls of matter-light interactions~\cite{Guery-Odelin2011}. Among various trapped cold atomic species, strontium (Sr) from group II alkaline-earth element has attracted a lot of interest to realize various experimental platforms from quantum sensing and metrology~\cite{Katori2003,Ye2008,Derevianko2011} to quantum simulation of condensed matter physic problems with ultracold fermionic and bosonic gases~\cite{Dalibard2011,Bloch2012,Cooper2019}.

Sr with its two valence electrons exhibits ground state transitions with very different oscillator strengths. Apart from the dipole-allowed $^1$S$_0\rightarrow\,^1$P$_1$ transition that offers large scattering rate (linewidth 32~MHz) with large momentum transfer, typically used in first stage magneto-optical trap (MOT), the intercombination $^1$S$_0\rightarrow\,^3$P$_1$ transition (linewidth 7.5~kHz) is ideally suited for narrow line cooling in a second MOT stage, reaching a temperature near the recoil limit~\cite{Xu2003,Loftus2004,chaneliere2008three,chalony2011doppler}. Another one of great importance is the doubly forbidden $^1$S$_0\rightarrow\,^3$P$_0$ clock transition, which is weakly allowed in $^{87}$Sr due to the hyperfine interaction through nuclear spins~\cite{Marty2007}, whereas it can can be magnetically induced in bosonic isotopes, where the nuclear spin is zero~\cite{Taichenachev2006,Santra2005,Hong2005,magic_B,Katori_magic_circular}. This ultra-narrow clock transition is a promising candidate for a new frequency standard with redefinition of the second~\cite{Akatsuka2008,Letargat2013,Ludlow_clock,Bothwell2019}. The clock transition is also useful for atom interferometers involving internal states with a large energy difference~\cite{Tino_interferometer, graham2013}, quantum simulation of many-body spin-orbit coupling physics~\cite{Kolkowitz_Qsimulation}, non equilibrium phenomena~\cite{Rajagopal_Qsimulation}, and quantum computation~\cite{Blatt_Qcomputation,PeterZoller_Qcomputation,Lukin_Qcomputatiom}.} 

In most of these applications, a compact and high-flux source of cold atoms is an important requirement. Due to its low vapor pressure at room temperature, Sr has to be heated to a high temperature in an oven to achieve a sufficient vapor pressure. This high temperature leads to a broad atomic velocity distribution and requires a bulky Zeeman slower. In addition, state-of-the-art sources of cold atoms, are often equipped with a 2D-MOT to transversely compress and deflect the beam toward a high vacuum main chamber~\cite{yang2015high}.  Using conventional Zeeman slowers and 2D-MOTs with electromagnets require complex designs with substantial energy consumption. To address these issues, an alternative approach with permanent magnets has been successfully implemented using Li~\cite{Walraven2009Li},  Na~\cite{lamporesi2013compact} and Sr~\cite{Widemuller_Sr,barbiero2020} in a hybrid configuration where the Zeeman slower and the 2D-MOT share the same magnets configuration, reducing the system size. In addition, the loading of the 2D-MOT occurs along a transverse direction making the hot atomic source naturally away from a direct line of sight to the region where the experiments are carried out.

In this article, we first present a compact ultracold Sr experimental system that uses the transverse loading of atoms into the 2D-MOT, and study its performance on $^{88}$Sr. To improve the atomic flux, we implement a hybrid bi-color slowing scheme by introducing an additional beam that is co-propagating and cross polarized with respect to the usual Zeeman slowing beam. Overall, a 11-fold increase in the cold atoms number is achieved when the hybrid slower is turned on.

In a second part, we address the issue of stray magnetic field fluctuations that affect the ultracold cloud position when the MOT is operated on the narrow linewidth $^1$S$_0\rightarrow\,^3$P$_1$ intercombination line at 689~nm. In addition, a control of the magnetic field environment is important for applications relying on precise transition frequency measurements.   The sources of stray magnetic fields in our lab include the Earth's magnetic field, the 50 Hz magnetic field radiated by the AC power lines, and other magnetic field noise sources due to human activities. For the latter, random jumps of tens of mG may occur during the experiment, affecting the reproducibility of the 689~nm MOT. A suppression of the stray fields is highly desirable. There are two main approaches, namely a passive shielding or an active compensation. Passive shielding with high-magnetic-permeability materials can reduce the DC noise by up to six orders of magnitude, down to a level of a few $\upmu$G~\cite{Magnetic_shield}. However, the shields are less effective at 50~Hz and higher frequencies~\cite{Dedman}. Moreover, it is challenging to properly shield a complex experimental setup where large bias magnetic fields and a large number of optical axes are required. On the other hand, active compensations using feedback loops have shown good performances reducing the stray magnetic field fluctuations below few hundreds of $\upmu$G, with a bandwidth above few hundreds of Hz~\cite{Ringot2001, botti2006, Dedman,Smith2011,Quieting_Bfield}. Here, we implement an active compensation system with two feedback loops, one for low frequency (< 5~Hz), and one centered on 50~Hz. Thanks to a pause-and-hold strategy, we compensate for stray fields even when the probes' readings are saturated by strong magnetic fields applied during the experiment sequences. 

This article is organized as follows: In Sec.~\ref{sec:ExptSetup}, we present a general overview of our experimental apparatus. The experimental realization and results of the bi-color slowing scheme on the broad $^{1}$S$_{0}\rightarrow\,^{1}$P$_{1}$ cooling transition are discussed in Sec.~\ref{sec:bicolor}. Then, we describe, in Sec.~\ref{sec:bsens}, the implementation of the active feedback compensation of magnetic fields, and demonstrate its ability to suppress the fluctuations in the MOT position due to magnetic field noise when operating on the narrow 689~nm $^{1}$S$_{0}\rightarrow\,^{3}$P$_{1}$ transition. Some concluding remarks are presented in Sec.~\ref{sec:conclusion}.

\section{Experimental setup}\label{sec:ExptSetup}

\begin{figure*}
\includegraphics[width=0.8\textwidth]{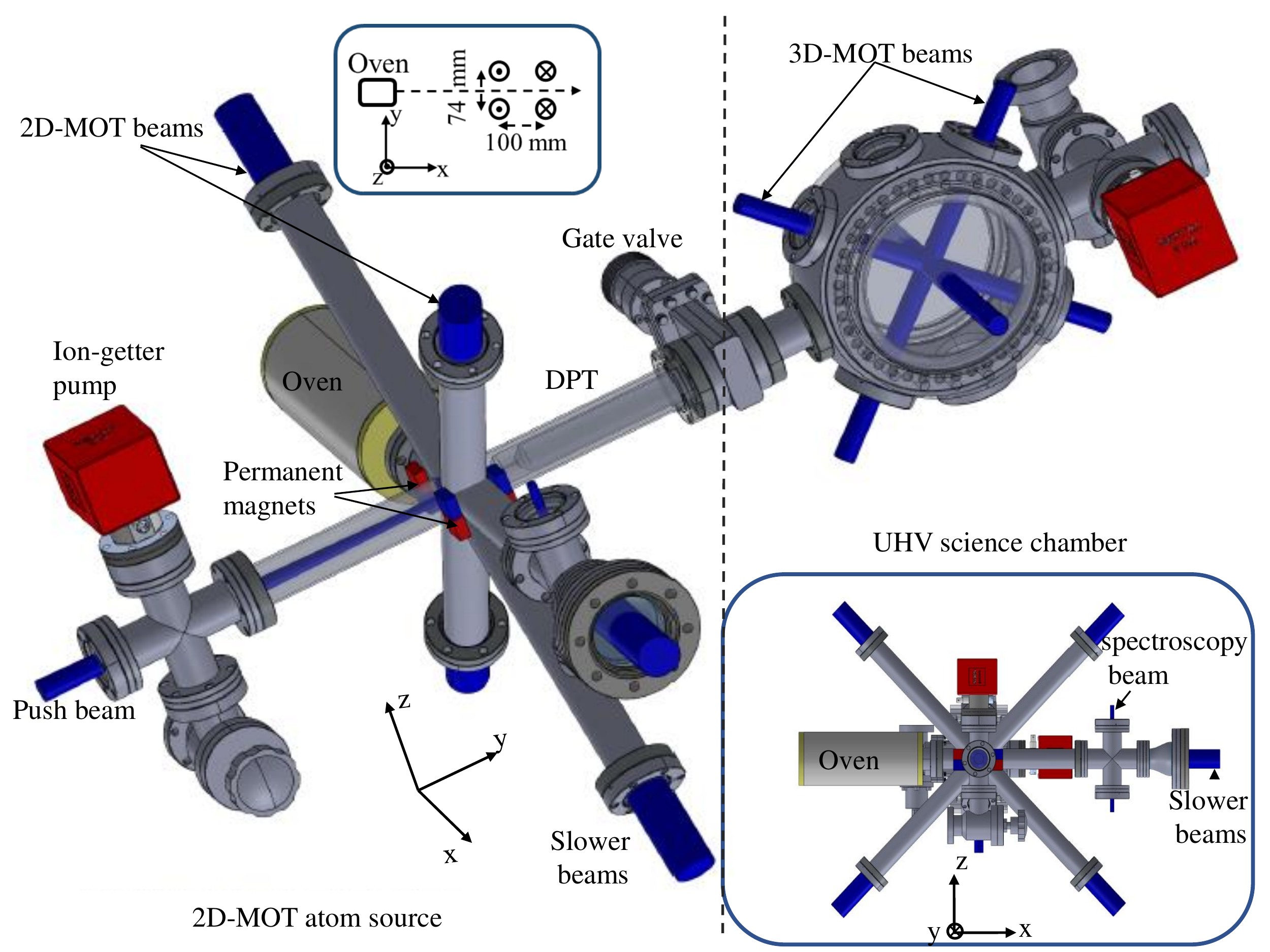}
\caption{A three-dimensional drawing of the experimental setup. The setup consists of two main parts, a 2D-MOT atomic beam source and a UHV science chamber, which are separated by a gate valve and a differential pumping tube (DPT). Four stacks of permanent magnets, color-coded blue and red to indicate the polarity, are arranged around the center of the 2D-MOT chamber to provide the magnetic field needed for both the 2D-MOT and Zeeman slower. The blue cylinders in the drawing represent the 461~nm laser beams used for 2D-MOT, 3D-MOT, bi-color slowing, pushing and spectroscopic reference. The bottom right figure shows an alternative view of the setup along the push beam direction. The inset on top shows a schematic of the orientation and position of the dipole moment generated by the permanent magnets. }\label{fig:fig1}
\end{figure*}

A sketch of the experimental setup is shown in \fig{fig1}. Our design consists of a dual vacuum chamber configuration, similar to Refs.~\cite{lamporesi2013compact,Widemuller_Sr,barbiero2020}. 
The two main parts of the setup include a cold atomic beam production stage, and an ultrahigh vacuum (UHV) science chamber where ultracold atoms are produced in a MOT operating at $461$~nm, followed by $689$~nm. The cold atomic beam stage is composed of an oven, a compact Zeeman slower and a 2D-MOT. The 2D-MOT is produced inside an eight-way stainless steel cross of 40 mm outer diameter tubing with a zero magnetic field line along the $y$-axis. Four arms of the eight-way cross are positioned in the vertical plane, at $45\degree$ with respect to the $x$-axis  and sealed with vacuum viewports to provide optical access for the 2D-MOT beams. Each arm has a length of 30~cm to avoid the deposition of thermal Sr atoms on the glass windows. The cold atomic beam stage is connected to the UHV science chamber along the $y$-axis, separated by a  23-mm-long differential pumping tube (DPT) with 2~mm inner diameter, and a gate valve. Both the atom source and science chamber on either side of the differential pumping tube are pumped by hybrid getter-ion pumps.

The 2D-MOT is loaded transversely by an effusive atomic beam along the $x$-axis produced by an oven mounted horizontally to the eight-way cross. The oven is located approximately 14~cm away from the 2D-MOT center, and is made of a stainless steel cylinder filled with a few grams of natural abundance Sr granules. In the usual operation, the oven is heated to about 520$\degreecelsius$ to attain an optimum Sr vapor pressure. The oven temperature is stabilized to ensure a constant flux of the atomic beam. A square array of about 900 Monel 400  micro-tubes forms the oven nozzle. Each of these micro-tubes are 1-cm-long with inner and outer diameters of 0.2 and 0.4~mm, respectively. The nozzle is heated at around 50$\degreecelsius$ above the oven temperature to prevent clogging by Sr deposits~\cite{yang2015high}. We installed a double layer heat insulation and water cooling plates around the oven to thermally isolate it from the external environment and the other parts of the vacuum setup.

The 2D-MOT is realized on the $461\,$nm dipole-allowed $^1$S$_0\rightarrow\,^1$P$_1$ transition (linewidth $\Gamma_b/2\pi=32\,$MHz) using two orthogonal pairs of retro-reflected laser beams that have opposite circular polarization. The beams are frequency red-detuned from the transition by $28\,$MHz~$=0.9\Gamma_b/2\pi$, with a total power of $P=110\,$mW and a beam waist of $1.25\,$cm. It corresponds to a total intensity of $88\,$mW cm$^{-2}\simeq 2.1I_s$, where $I_s=42$~mW/cm$^{2}$ is the saturation intensity of the transition. The beams intersect at the center of the eight way cross (see \fig{fig1}). The 2D quadrupole magnetic field used to trap the atoms in the $xz$-plane is produced by four stacks of NdFeB permanent magnets (N750-RB from
Eclipse Magnetic Ltd.) arranged around the center of the eight-way cross.

Atoms cooled in the 2D-MOT are pushed toward the science chamber with a longitudinal velocity of $\sim$15~m/s by a weak 461~nm near resonant push beam (detuning: $-2\,$MHz, power $0.05\,$mW). The push beam is slightly focused into the 2D-MOT, and is defocused at the location of the 3D-MOT to minimize its mechanical action on the cold atoms in the 461~nm MOT. The 461~nm MOT is produced using three pairs of mutually orthogonal beams, a magnetic field gradient of 50~G/cm, and two repumper beams at 707 and 679~nm~\cite{yang2015high}. The MOT beams are frequency red-detuned from the transition by $58\,$MHz~$=1.8\Gamma_b/2\pi$, and have a total intensity of $I=1.07 I_s$.

The 461~nm laser beams are obtained by frequency doubling a 922~nm laser. Along the vertical direction of the four-way cross mounted opposite to the oven, we perform a frequency-modulation spectroscopy of the atomic beam to lock the 922~nm laser frequency.

\section{Bi-color slowing scheme}\label{sec:bicolor}

To improve the performance of the cold atomic beam source, we send a pair of 461~nm laser beams, counter-propagating with the atomic flux produced by the oven, to perform a hybrid bi-color atomic beam slower on the $^1$S$_0\rightarrow\,^1$P$_1$ transition of $^{88}$Sr.

The hybrid bi-color slowing scheme consists of a Zeeman slower that operates in combination with an additional beam that slows atoms via radiation pressure on the magnetically insensitive $m=0\rightarrow m=0$ transition of the 461~nm line. 

\begin{figure}
    \centering
    \includegraphics[width=0.45\textwidth]{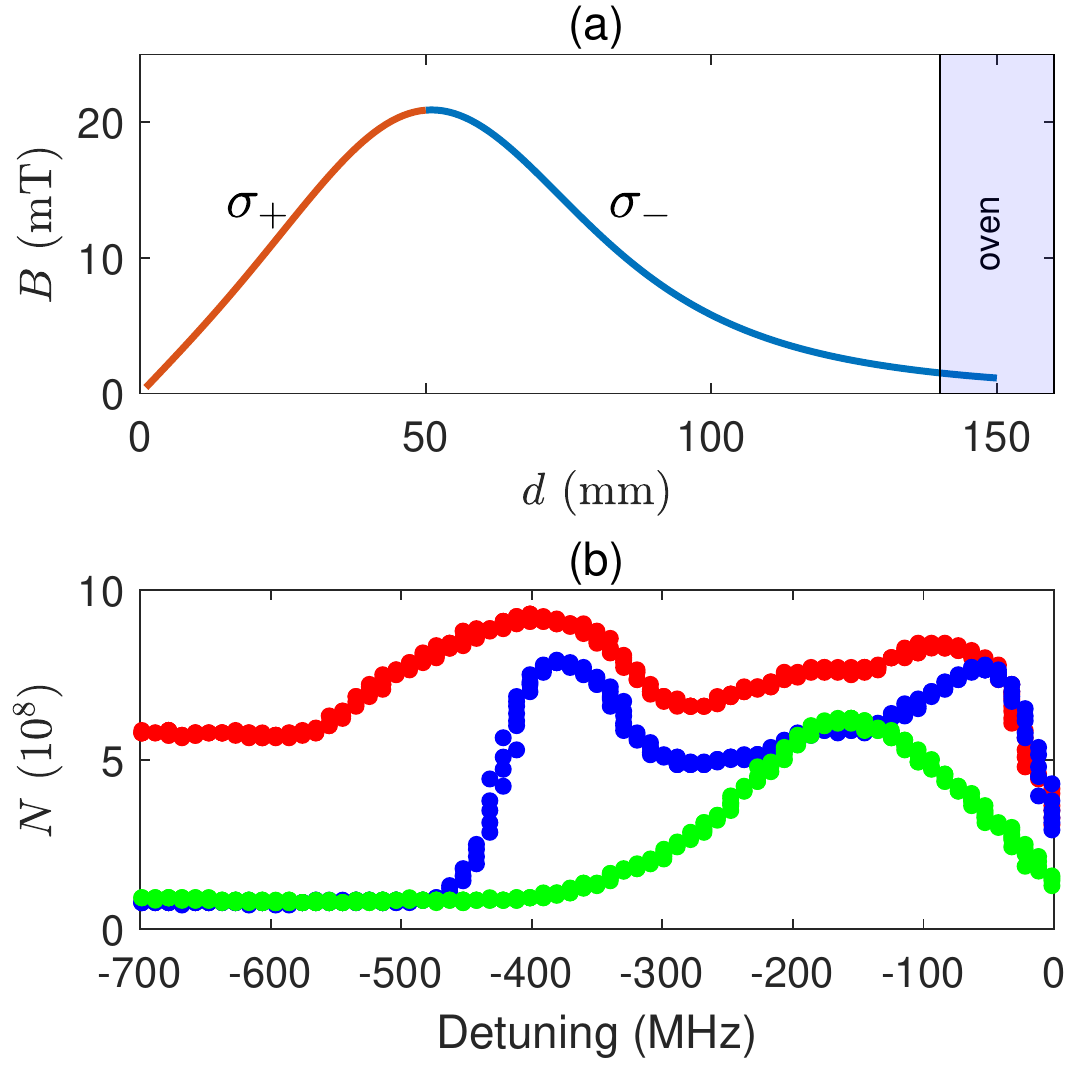}
    \caption{(a) Magnetic field $B$ of the permanent magnets as a function of distance $d$ from the 2D-MOT center along the $x$-axis. (b) Performance of the slowing scheme as measured by the total number of atoms $N$ loaded in the steady-state 461~nm MOT. The blue curve is obtained with the Zeeman slowing beam only (horizontally polarized beam). The green curve shows the effect of slowing on the $m=0\rightarrow m=0$ transition (vertically polarized beam). The red curve shows the performance of the hybrid scheme comprising both horizontally and vertically polarized beam. The vertically polarized beam is set to a frequency detuning of $-180$~MHz. In all cases, the total power is 160~mW}
    \label{fig:fig2}
\end{figure}

We first characterize the Zeeman slower of the hybrid scheme. The magnetic field profile from the permanent magnets along the $x$-axis, although not optimized, is still suitable to perform Zeeman slowing~\cite{Widemuller_Sr}. In the whole region of interest, the magnetic field direction is mainly along the z-axis. Its magnitude is plotted in~\fig{fig2}(a) as a function of $d$, the distance from the 2D-MOT center. We evaluate the performance of the slowing scheme by measuring the number of atoms trapped in the 461~nm MOT in the steady-state regime, which is achieved after 1.5~s of loading time. For a Zeeman slowing beam polarized along the $y$-axis, we find that the Zeeman slower reaches its peak efficiency at detunings of $\sim-380\,$ and $\sim-30\,$MHz [see the blue curve in~\fig{fig2}(b)]. Indeed, the horizontal linear polarization is decomposed into two circular polarization eigen-modes addressing the $m=0\rightarrow m =\pm1$ transitions, respectively. Each circular polarization component is associated with a Zeeman slowing effect on one of the magnetic field slope regions. At large negative frequency detuning of $\sim-380\,$MHz, we address the negative magnetic field slope [blue curve in~\fig{fig2}(a)], where the large Doppler effect of atoms at the oven output is not compensated by a substantial Zeeman shift. In this case, we need a negative Zeeman shift corresponding to the $m=0\rightarrow m =-1$ transition. We end up with the opposite configuration on the positive slope [red curve in~\fig{fig2}(a)]. Here, the deceleration starts with a large magnetic field bias and positive Zeeman shift to compensate for the larger positive Doppler shift of the Zeeman beam seen by the faster moving atoms. In previous works, only one of the frequencies is chosen to operate the Zeeman slower~\cite{Walraven2009Li,Widemuller_Sr}. Usually large negative frequency detunings are favored to limit the mechanical action of the Zeeman beam on the 2D-MOT. We note also that the orientation of the magnetic field means that only 50\% of the optical power is used for Zeeman slowing.

We now consider the slowing effect of a vertically polarized beam. In this case, the $m=0\rightarrow m=0$ magnetic-insensitive transition is used to decelerate the atoms. The performance with respect to the laser frequency detuning is indicated as the green curve in \fig{fig2}(b). Using only the vertically polarized beam, the most effective slowing is achieved at a frequency detuning about  -180~MHz. 

By fixing the frequency of the vertically polarized beam at -180~MHz, we scan the frequency of the horizontally polarized beam with the same total power to obtain the red curve in \fig{fig2}(b). With this hybrid bi-color slowing scheme, we improve the overall number of atoms in the 461~nm MOT for a detuning of $\sim-400\,$MHz, meaning slightly more red-detuned than the single-color Zeeman slower case. We obtain a total of 9$\times$10$^8$ atoms in the 461~nm MOT. In addition, we note that the optimal number of atoms occurs on a broader slowing beam frequency range, leading to a slowing scheme that is more robust to potential frequency fluctuations.

\begin{figure}
    \centering
    \includegraphics[width=0.5\textwidth]{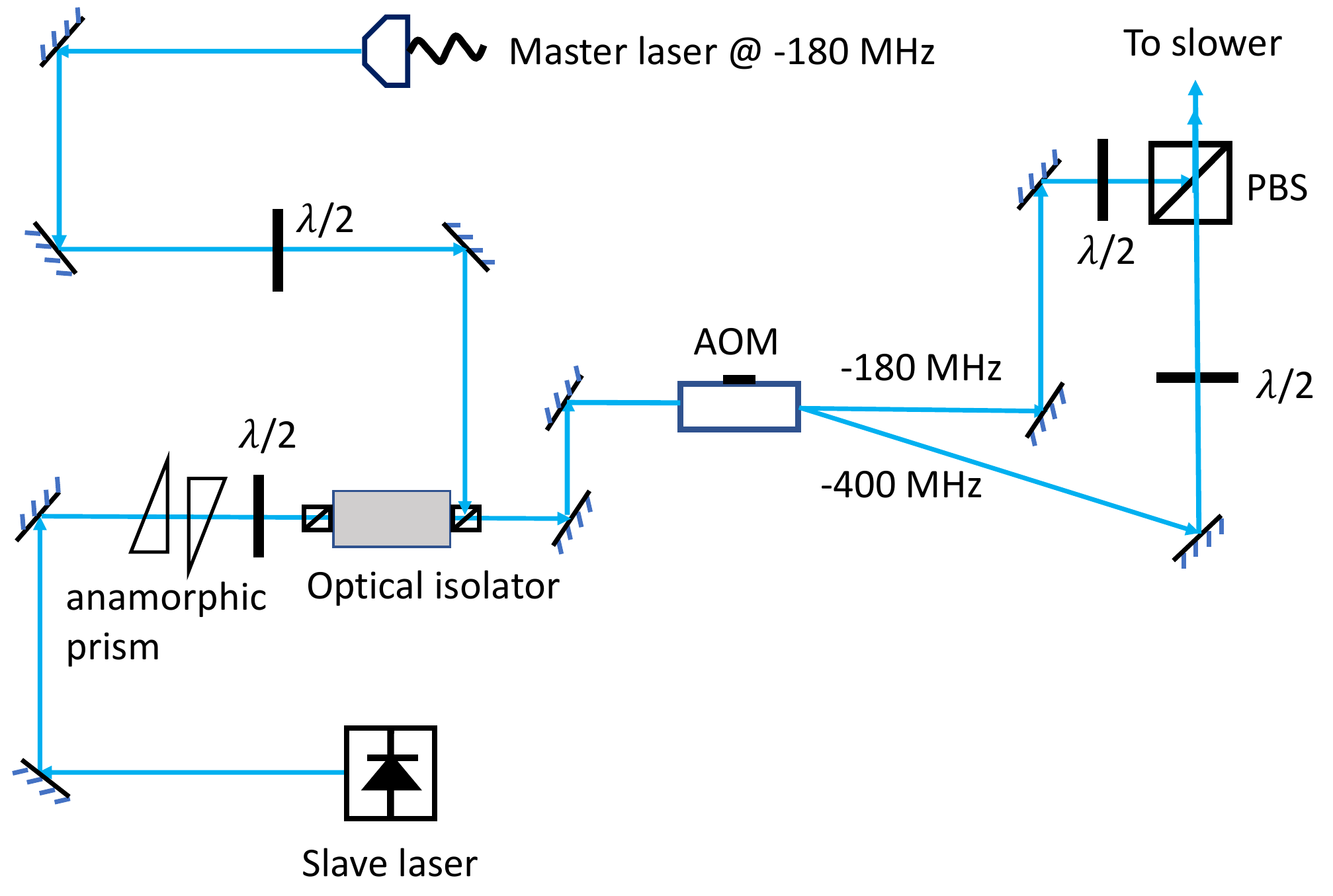}
    \caption{Optical setup to generate the two overlapped slowing beams for the hybrid bi-color slowing system.}
    \label{fig:Zeemanlaser}
\end{figure}

The hybrid bi-color slowing scheme is optimized using two independent laser systems. However, thanks to the cross-polarization between the two beams, we are able to implement a single laser source with efficient use of the laser power for daily operations as shown in \fig{Zeemanlaser}. We take 1.5~mW from the 461~nm laser source power to inject a 500~mW laser diode (Nichia, NDB4916). The seed laser beam used for injection locking is frequency shifted with an acousto-optic modulator (AOM) by -180~MHz from the resonance of the $^1$S$_0\rightarrow\,^1$P$_1$ transition. The laser output from the slave laser passes through another AOM at 220 MHz to provide the two slowing beams required in the hybrid bi-color slowing scheme. The zero-order of the AOM at a detuning of -180~MHz is utilized to address the $m=0\rightarrow m=0$ transition with the polarization parallel to the magnetic field; the minus-one-order at a detuning of -400~MHz is used to slow down the atoms via $m=0\rightarrow m=-1$ transition. The two diffraction orders of 0 and -1 are efficiently recombined  using a polarizing beamsplitter (PBS). The overlapped collimated beams with a diameter of about 25 mm are sent through a sapphire viewport, onto the oven.

\begin{figure}
    \centering
    \includegraphics[width=0.5\textwidth]{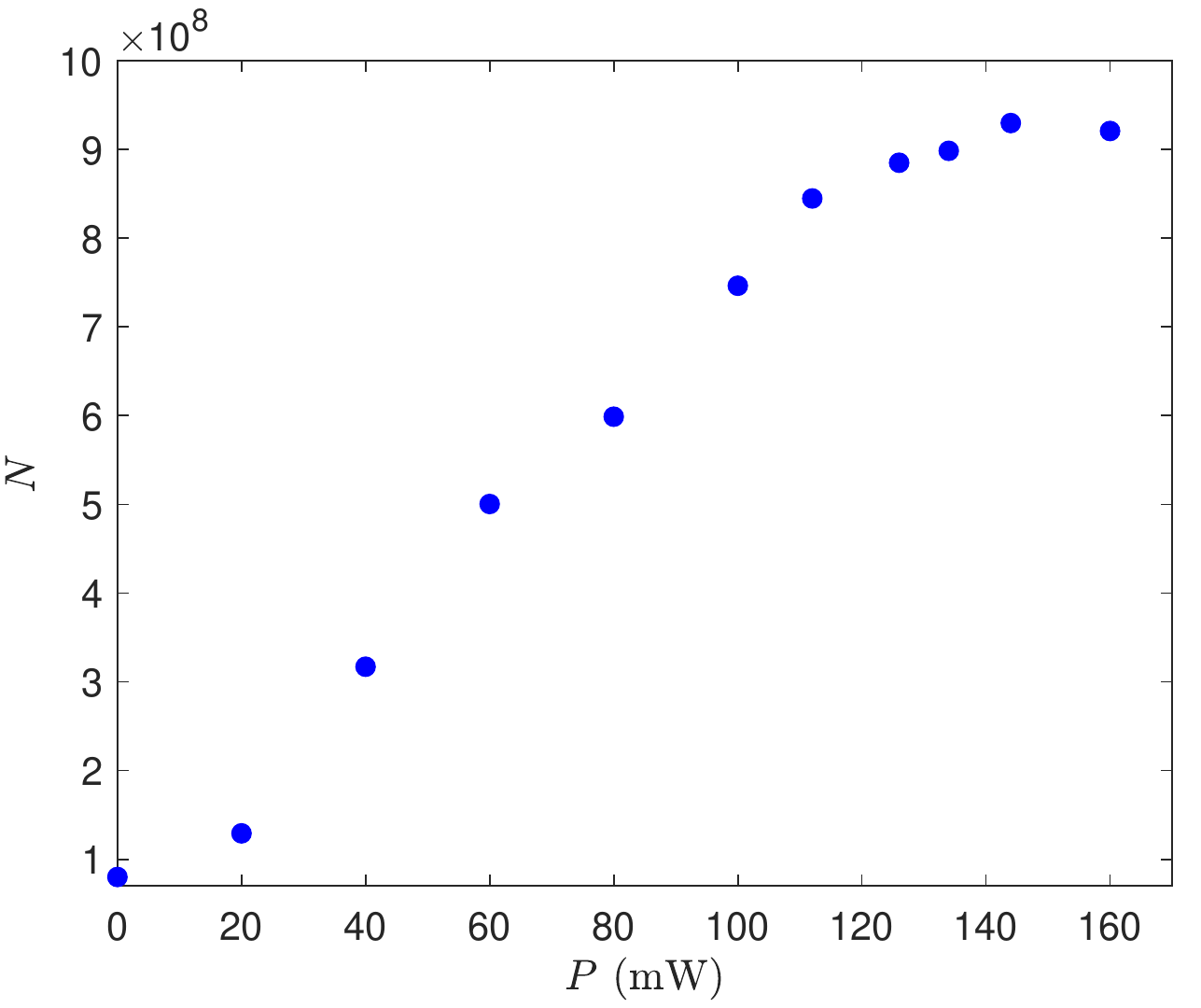}
    \caption{Number of atoms in 3D-MOT with respect to the condition when the hybrid slowing beams are turned off, plotted as a function of total power $P$. $P$ is divided equally between the two polarization components.}
    \label{fig:Zeemanpower}
\end{figure}

On Fig.~\ref{fig:Zeemanpower}, we show the cold atoms number $N$ as function of the slower total power $P$. When the optical beam is off, $N\sim 10^8$, indicating a substantial trapping efficiency of the bare atomic beam by the 2D-MOT. We note that, as expected, this value is also obtained with a far off-resonant slower (see green and blue data points in Fig.~\ref{fig:fig2} for detuning beyond -500~MHz). For a slower total power above 140~mW, the performance of the bi-color slowing scheme saturates, so no further increasing of the optical power is necessary. In practice, we use $P=\,$160~mW to minimize the atoms number shot-to-shot fluctuations in the MOT due to the drift in the optical power of the slowing beams. Overall, the bi-color slowing scheme enhances the atoms number by one order of magnitude leading to $N\sim 10^9$, a value similar to the current state-of-art setups \cite{yang2015high}, but obtained here with a more compact and simple system.

\section{Magnetic field compensation}\label{sec:bsens}
The narrow $^1$S$_0\rightarrow\,^3$P$_1$ intercombination line transition is sensitive to magnetic fields at the mG level \cite{pandey2016linear}. Due to human activities, we observe occasional jumps of the ambient magnetic field on the order of 10~mG.  We implement an active control system to compensate the external magnetic field, including the random jumps, other slow varying and DC contributions, as well as the AC 50~Hz line contribution. In the following, we describe the main components of this system. A more extended description can be found in Ref.~\cite{kwong2017}.

\begin{figure}
    \centering
\includegraphics[width=0.5\textwidth]{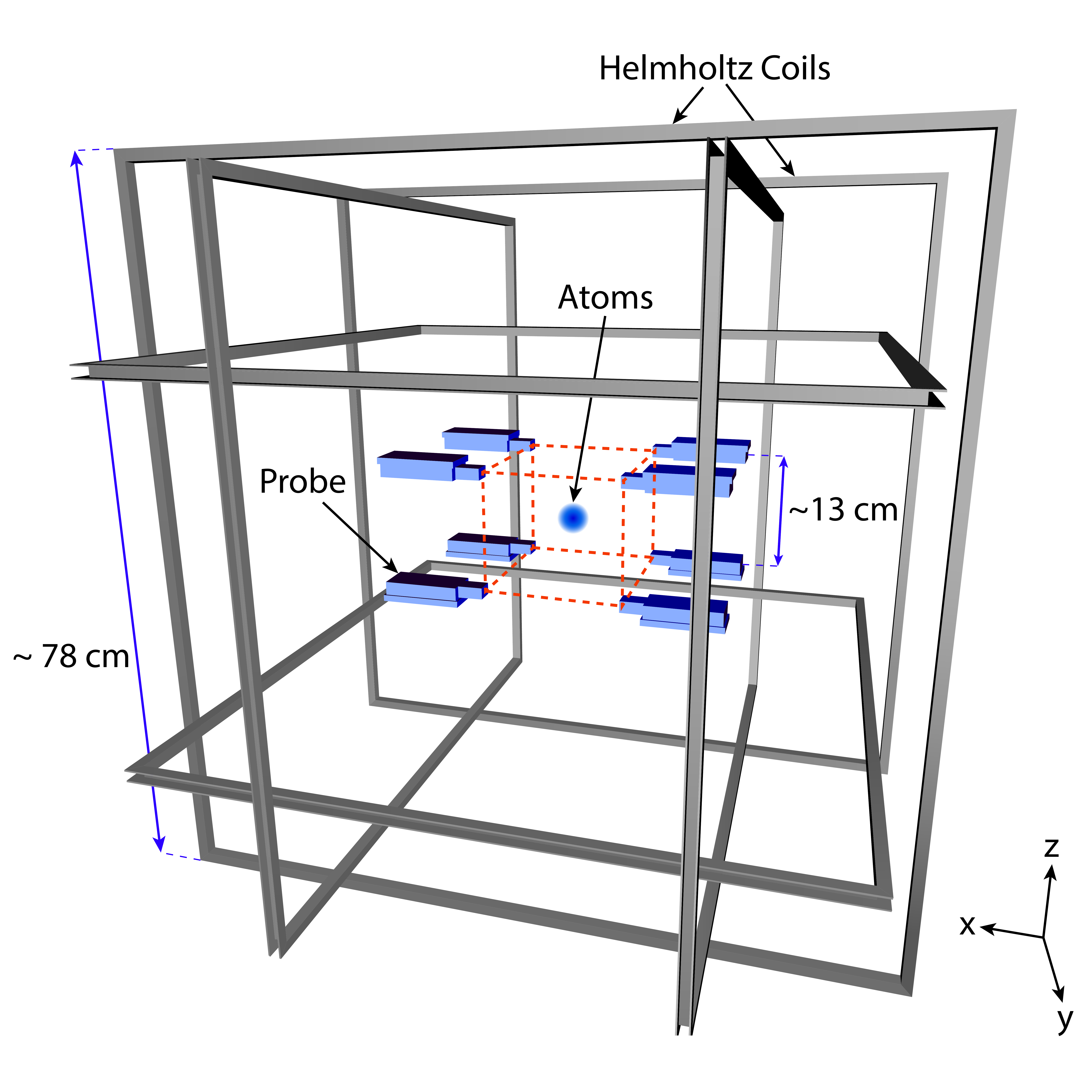}
    \caption{Arrangement of the eight magnetic-field probes and the compensation coil system centered on the ultracold atomic cloud.}
    \label{fig:compensation}
\end{figure}

Our active control system includes three main components: a probe network consisting of eight three-axis magnetic-field probes arranged in a cuboid geometry centered on the ultracold atomic cloud position, a personal computer (PC) for digital signal processing, and three pairs of quasi-Helmholtz coils to perform the magnetic-field active compensation in 3D. The probe network and coils are shown in \fig{compensation}.

\subsection{Magnetic probe}
Each three-axis magnetic-field probe comprises one Honeywell one-axis (HMC1001) sensor and one two-axis (HMC1002) sensor. These sensors are anisotropic magnetoresistive sensors~\cite{Kwiatkowski1986, bertoldi2005, bertoldi2006}. Each axis of the HMC sensor contains four permalloy resistive strips arranged in a Wheatstone bridge configuration. The sensor operates in the linear regime of the magnetoresistive response where the output bridge voltage is a linear function of the magnetic field. The sensitivity of the sensor relies on the alignment of the permalloy magnetization along an easy axis. A strong external magnetic field could disrupt the permalloy magnetization, resulting in a loss of sensitivity. To maintain the sensors' sensitivity, before each measurement we align the resistive strip magnetization by means of a current pulse sent in the set/reset strap of the chip. Furthermore, we reverse the magnetization direction to have it parallel or antiparallel to a chosen axis in successive measurements. Taking the difference of two such measurements removes magnetic-independent offsets and their temperature dependence~\cite{bertoldi2006}. 

To achieve an optimum sensitivity, the bridge output voltage for each axis is  amplified with a gain of 200, and digitized with a 16-bit analog-to-digital converter (ADC).  We achieve a resolution of around 34~$\upmu$G per digital level for each axis of the probe, and an operational range of approximately $\pm1$~G. 

We characterize the noise spectral density of the three-axis magnetic probes in a zero Gauss chamber (ZGC). The noise spectral density of each probe is typically flat between 100~mHz and 50~Hz, with a value of around $2.7$~$\upmu$G/$\sqrt{\text{Hz}}$. This is consistent with the Johnson noise of an 850~$\Omega$ resistor (the typical bridge resistance of the HMC sensors) after taking into account the effect of the op-amp amplification and the ADC. The root-mean-square (rms) noise calculated for a bandwidth up to the Nyquist frequency of 244~Hz is below 38~$\upmu$G, setting the limit on the measurement resolution.

\begin{figure}
    \centering
    \includegraphics[width=0.5\textwidth]{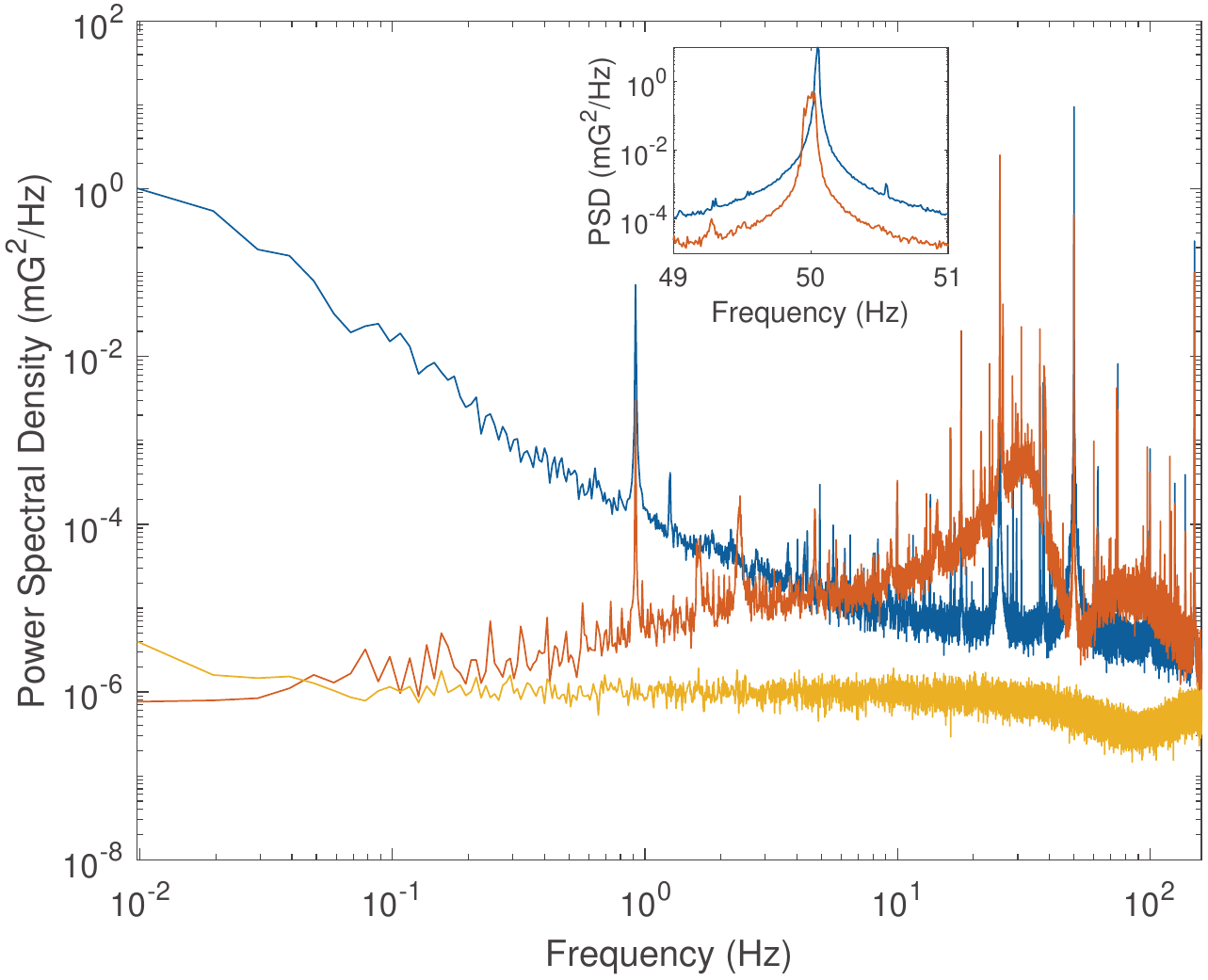}
    \caption{PSD of the magnetic field measurement along the $z$ direction. The blue (red) trace is without (with) active feedback control. The yellow trace is the spectrum of the average readings from all eight probes measured independently in the ZGC. The inset shows a zoom of the PSD around 50~Hz.}
    \label{fig:psd}
\end{figure}

\subsection{Probe network}\label{sec:probe}

Eight three-axis probes are arranged in a cuboid configuration around the science chamber with separations of 155, 115, and 123~mm along $x$, $y$ and $z$ axes, respectively. The probe readout time is synchronized across all the eight probes by a microcontroller. The probes take turns to send their readings to the microcontroller through a controller area network (CAN) protocol. A USB-CAN adapter is implemented to send all eight readings to a PC. The eight readings are then averaged to interpolate for the value of the magnetic field at the location of the atoms. The communication protocol limits our maximum sampling frequency for the probe network to 488~Hz. The Nyquist frequency of 244~Hz is sufficient to capture the relevant frequencies of the stray magnetic field. 

With the probe network mounted around our experimental setup, we perform a measurement of the Cartesian components of the magnetic field homogeneous contribution, by averaging the reading of the eight sensors. The power spectral density (PSD) of the magnetic field component along the $z$-axis corresponds to the blue trace in Fig.~\ref{fig:psd}. Similar spectra are obtained for the two remaining components along the $x$- and $y$ axes. For comparison, the yellow trace is the averaged noise density of the readings from the eight probes measured independently in the ZGC. In this uncorrelated noise condition, we find a noise floor of $\sim 1~\mu$G/$\sqrt{\text{Hz}}$. When installed in the lab the noise level at frequency $\gtrsim 5$~Hz is higher than the uncorrelated noise floor, possibly due to common mode noise from sharing the same power supply. On top of this, the probe network picks up stray magnetic field contributions at frequency $\lesssim5$~Hz, and AC noise components mainly at 50~Hz. More precisely, we find an \textit{rms} magnetic field noise of 150$\,\mu$G at low frequency (0.001~Hz to 5~Hz) and 510~$\mu$G between 10 and 100~Hz, dominated by the AC-line contribution.

\subsection{Feedback control}
We use three pairs of square coils in quasi-Helmholtz configuration along the three orthogonal directions ($x$, $y$, $z$) to correct the stray magnetic field. Each coil consists of 18 turns and is wound on a square aluminum frame with a side of about 78~cm. Compared to the size of the probe network, this gives a sufficiently good field uniformity for the probe network to accurately interpolate the magnetic field. Each pair of the compensation coils is connected to a low noise current source, and generates a magnetic field of around 370~mG/A.

We implement digital filters, and generate an error signal to perform a proportional-integral-derivative (PID) feedback control of the stray magnetic field. For the compensation of the frequency noise below 5~Hz, a low pass filter with cutoff frequency 10~Hz is implemented on the averaged probe readings before the PID is applied. 

To cancel the 50~Hz noise component, we first extract the quadrature amplitudes of the 50~Hz signal by mixing the probe reading with a 50~Hz signal, and then perform a low pass filtering with a cutoff frequency of 5~Hz. We then perform a PID control on the quadrature amplitudes before regenerating the feedback signal at 50~Hz. We allow for the 50~Hz phase of the feedback on the $x$, $y$ and $z$ components to be varied independently. We finally sum up both the low-frequency and 50~Hz feedback signals.

\subsection{Performance of the active control system}

Beforehand, the performance of the active control system is evaluated when the experimental sequence is not running. The PSD is shown in Fig.~\ref{fig:psd} as the red trace. Compared to the blue trace we see that the PID feedback control is able to substantially reduce the low frequency components of the magnetic field noise up to 5~Hz. We observe a reduction in the PSD by more than five orders of magnitude close to DC, and one order of magnitude at 50 Hz (see inset of Fig.~\ref{fig:psd}). We note that the 50~Hz feedback loop leads to a small shift and broadening of the 50~Hz peak, which we could not explained. An increase in the PSD due to the control of the 50 Hz component is also observed for frequencies above 10~Hz, with respect to the blue curve reducing the performance of the feedback loop. With the active control system, we find a typical \textit{rms} magnetic field noise of 10~$\mu$G at low frequency (0.001 to 5~Hz) and  370~$\mu$G between 10 and 100~Hz. In both frequency range we observe a reduction of the noise compared to the open-loop values (see Sec. \ref{sec:probe}).

\begin{figure}
    \centering
    \includegraphics[width=0.5\textwidth]{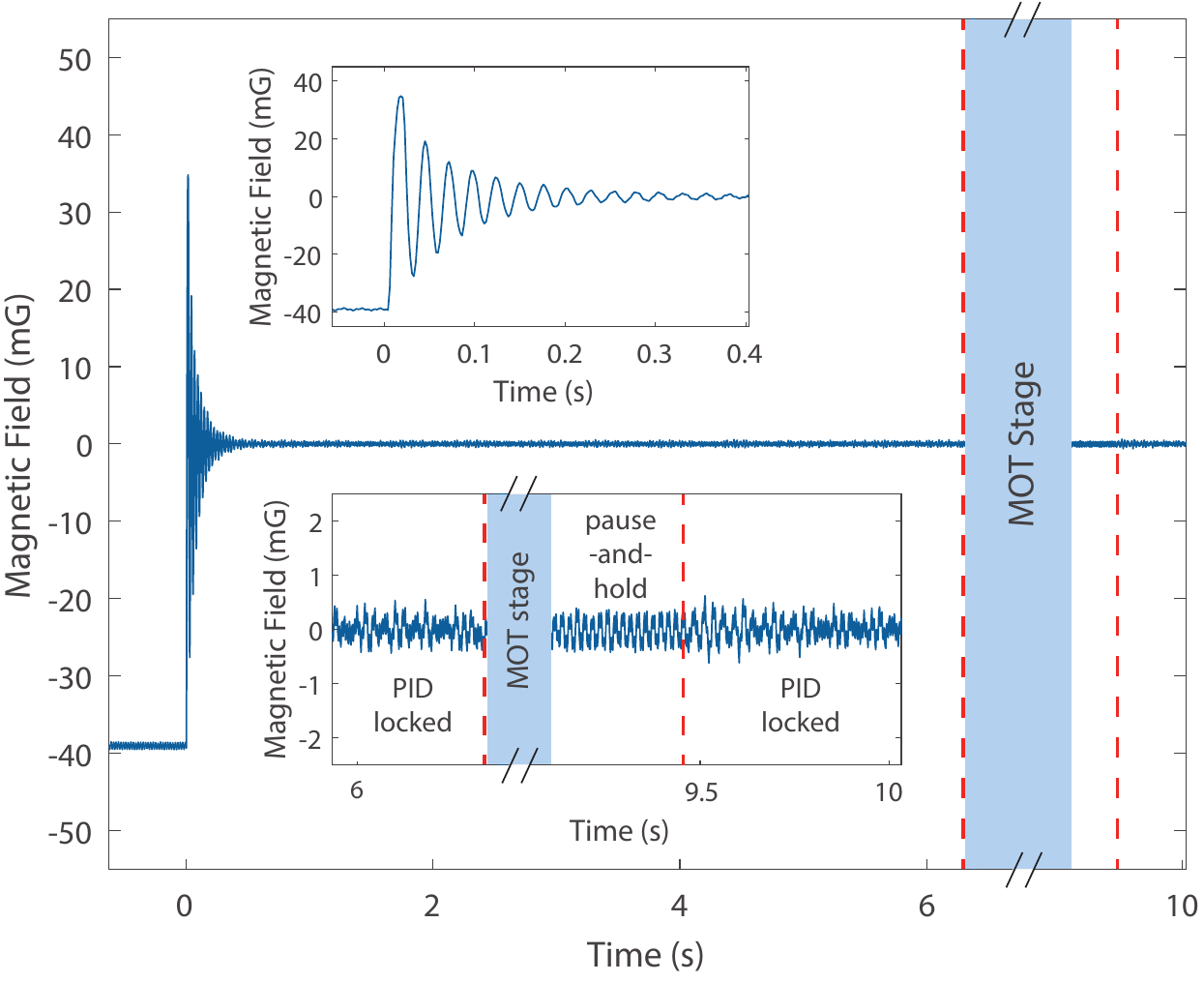}
    \caption{Magnetic field measurement during the experimental sequence. The top inset zooms into the magnetic field when the compensation is applied. The bottom inset zooms into the magnetic field during different stages of the experimental sequence. The pause-and-hold stage occurs between the two vertical red dashed lines.}
    \label{fig:bfieldexp}
\end{figure}

The operational range of the probes means that their readings become saturated whenever a magnetic field larger than 1~G is applied, which occurs for instance during the MOT stages. We apply a pause-and-hold strategy to compensate for the stray field during these stages, by retaining the feedback values before the application of the magnetic field. An example of the typical magnetic field reading during an experimental sequence is shown in Fig.~\ref{fig:bfieldexp}. Here, the initial magnetic field offset of 40~mG is larger than typical jumps experienced in the lab. The $1/e$ settling time, as shown in the upper inset, is about 100~ms. Thus, a time interval larger than 100~ms is required between each pause-and-hold stage to ensure that the magnetic field compensation is correctly applied. During the typical pause-and-hold stage, the magnetic field fluctuations remain less than $\pm500\,\mu$G, with an rms value of 210~$\mu$G over the measurement bandwidth.

\begin{figure}
    \includegraphics[width=\linewidth]{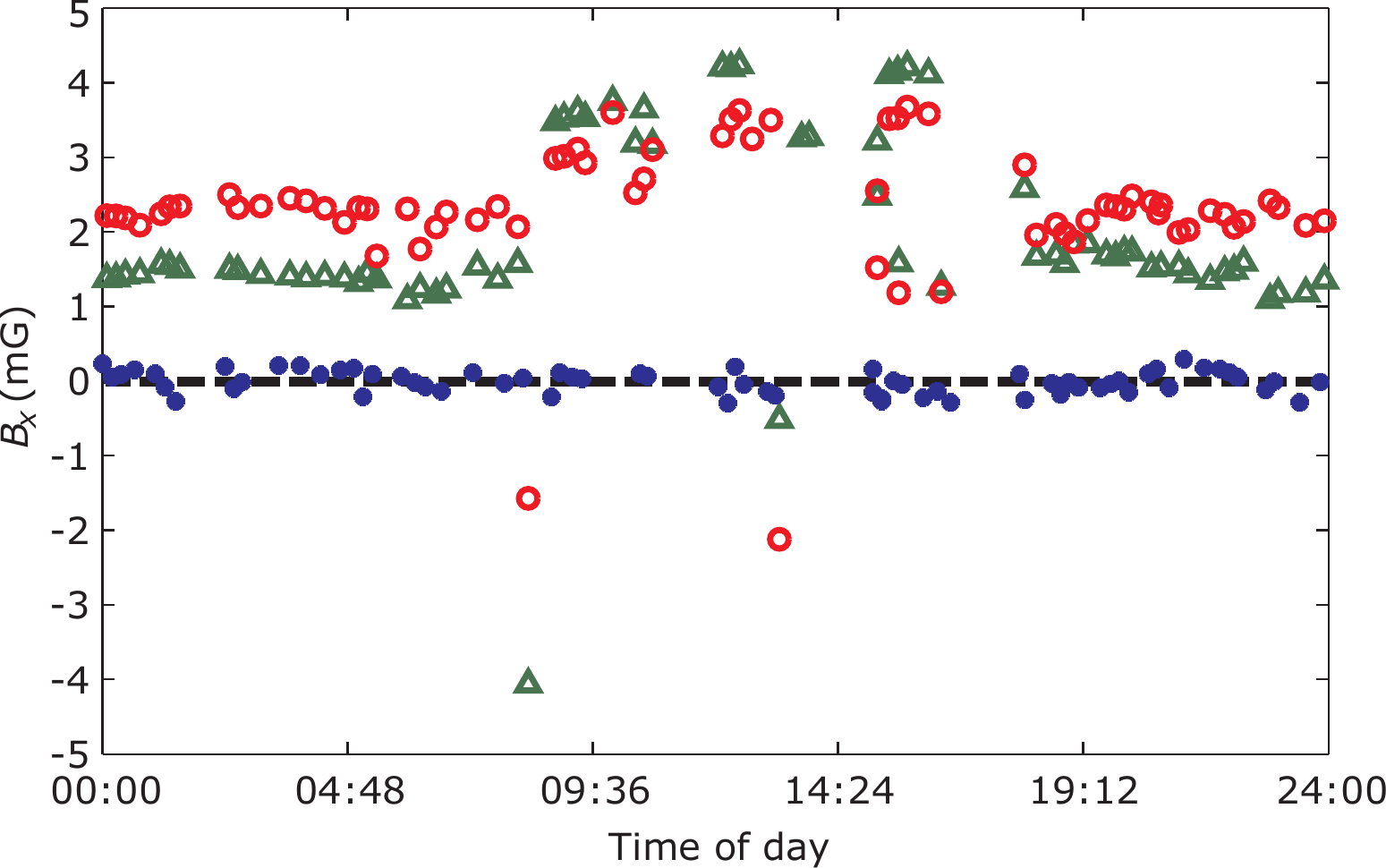}
  \caption{Measurement of the magnetic field variation in the course of 24 h. The green open triangles correspond to the sensors reading of the magnetic field along the $x$-axis. The red open circles and the blue circles are the reconstructed values of the magnetic field thanks to Faraday rotation measurement for unlocked case and locked to zero magnetic field (dashed line) case, respectively.}
  \label{fig:faraday}
\end{figure}

The green open triangles in Fig.~\ref{fig:faraday} show the typical magnetic field fluctuations measured in the lab within 24 h. From 9~a.m. to 7~p.m., we observe larger fluctuations due to nearby human activities. The magnetic field values are extracted from the sensors reading with an open feedback loop. Once the feedback loop is closed the magnetic field reading goes to zero (dashed lines) with fluctuations not visible for the range used in Fig. \ref{fig:faraday}. We perform a measurements of the magnetic field using Faraday rotation on the cold atomic gas. At low laser intensity, the Faraday rotation angle $\theta$ depends linearly of the magnetic field, with a slope of $\Delta\theta/\Delta B\sim b_0\times 0.28\,$rad/mG, if the measurement is performed on the intercombination line \cite{pandey2016linear}. $b_0$ is the optical thickness measured at resonance and at zero temperature \cite{PhysRevLett.113.223601}.

The Faraday measurement is performed by sending a resonant probe on the intercombination line along the $x$-axis with a polarization at $\pi/4$ angle with respect to the $z$-axis. The polarization rotation is analyzed thanks to a polarizing cube with $s$ and $p$ channels along the $z$- and the $y$-axis, respectively. The channel imbalance is recorded on a CCD camera (see Ref.~\cite{pandey2016linear} for more details). The red open circles and the blue circles in Fig.~\ref{fig:faraday} correspond to the magnetic field extracted from the Faraday angle measurements in the unlocked and locked situation, respectively. For the unlocked case, we observe a qualitative agreement with the sensors reading. Mismatches are likely due to calibration bias of the Faraday response slope and overlooked nonlinear responses. Importantly, we extract a magnetic field value of $B_x=10(150)\,\mu$G when the feedback loop is closed, in quantitative agreement with the expected null value. 

\begin{figure}
    \includegraphics[width=\linewidth]{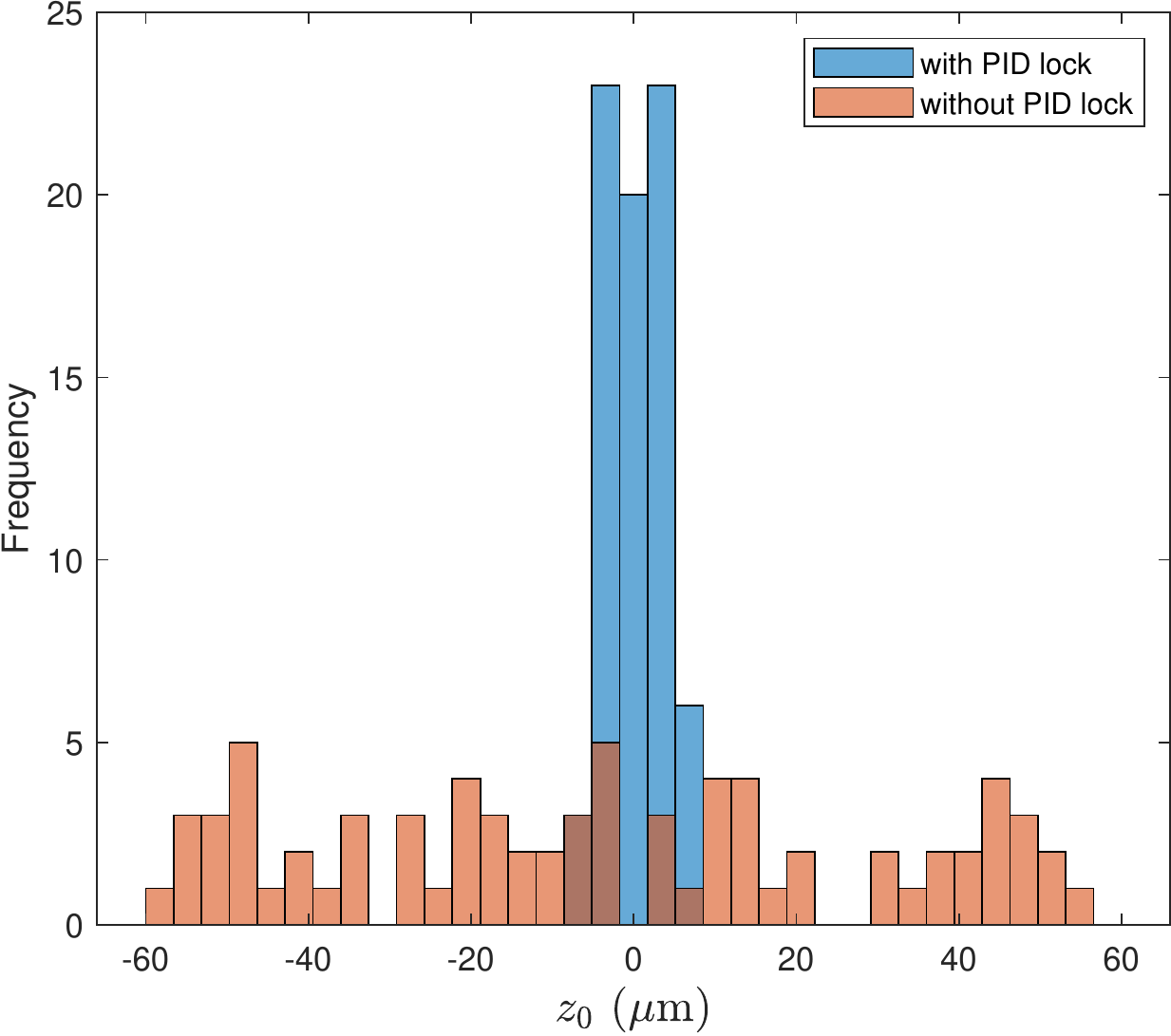}
  \caption{A histogram of the 689~nm MOT positions, $z_0$, along the vertical direction that is measured using fluorescence imaging, where blue (orange) bars shows the distribution of $z_0$ with the PID feedback loop activated (deactivated). For each of the two sets the measurement has been repeated 70 times. }
  \label{fig:fig6}
\end{figure}

As a final test of the performance of the active control system, we measured the position of the ultracold cloud in the $689\,$nm MOT along the $z$-direction with the PID feedback loop open or closed. The 689~nm MOT is produced following Ref.~\cite{yang2015high}. We apply random magnetic field jumps of 7.5~mG peak-to-peak to mimic the random magnetic field jumps due to human activities. Due to gravity, the $z$-position of the 689~nm MOT is shifted by 16~$\mu$m/mG. For each experimental run we apply a fix random value during the stage when the loop is closed, in contrast to the reality where magnetic field jumps can occur anytime. When a magnetic field jump occurs during the pause-and-hold stage, the active control system will compensate for it in the next PID locked stage. A comparison of the position of the 689~nm MOT subjected to this magnetic field fluctuation is shown in Fig.~\ref{fig:fig6}. We perform two experiments, one in open loop (orange bars) and another one in closed loop (blue bars). Without feedback, the random magnetic field modulation leads to a peak-to-peak variation in the MOT position of $\pm$60~$\mu$m. The active control system is successful in reducing this variation to less than $\pm7$~$\mu$m, a level that is consistent with the magnetic field fluctuations of less than $\pm0.5$~mG observed during the pause-and-hold stage in Fig.~\ref{fig:psd}.  This leads to a more consistent preparation of the cold atomic cloud.

\subsection{Addressing magnetic field gradients}

As discussed in Sec.~\ref{sec:ExptSetup}, the bi-color slowing scheme and the 2D-MOT are generated thanks to permanent magnets. The contribution of those magnets at the level of the 3D-MOT consists of  two gradient terms $\partial B_{x,z}/\partial z,x$, with an amplitude in the mG/cm range. Those gradients are too weak to disturb the cold gas in the MOT operating on the intercombination line, but can be a limitation for sensing applications requiring interferometers with spatially-extended surface area~\cite{hobson2021bespoke}. 

The sensor network measures the three Cartesian components of the magnetic field at eight positions around the MOT. Thus, we can extract the full magnetic field Jacobian matrix at the atomic cloud position, if the source is far enough from the sensors location to provide a good linear expansion of the magnetic field. This is indeed the case for permanent magnets of the  bi-color slowing scheme and the 2D-MOT.

The compensation of the off-diagonal components of the magnetic field Jacobian matrix can be done using Golay coils as developed for magnetic resonance imaging devices~\cite{hidalgo2010theory}. We note that since the magnetic field gradients should be stationary, no active compensation is necessary here.

\section{Conclusion}\label{sec:conclusion}

We propose a new hybrid bi-color atomic beam slower based on a compact design where the magnetic profile is imposed by the quadrupole magnetic field of a 2D-MOT. We use two crossed-polarized beams addressing the magnetic-sensitive $m=0\rightarrow m=-1$ transition and the magnetic-insensitive $m=0\rightarrow m=0$ transition of the $^1$S$_0\rightarrow\,^1$P$_1$ line of $^{88}$Sr. We observe an 11-fold improvement of the atoms number in the MOT when the bi-color slowing beam is turned on. This improvement is smaller than the 20-fold improvement of a Zeeman slower adopting an optimal magnetic field profile~\cite{yang2015high}. However, our system, featuring a short distance between the oven and 2D MOT, has a substantial loading of $\sim 10^8$ atoms without the slowing beams. As a result, the number of atoms is comparable in both setups when the slowing beam is on, namely $\sim 10^9$ atoms. Our system is hence characterized by similar performances in terms of atomic flux, but has the advantage of simplicity, compactness, and lower power consumption.

In addition, we propose an active control of the stray magnetic field, which can compensate fluctuations below $5\,$Hz, and around the $50\,$Hz of the AC line. We find a typical rms magnetic field noise of 10~$\mu$G at low frequency (0.001 to 5~Hz) and  370~$\mu$G between 10 and 100~Hz. With such an active compensation, we strongly reduce the magnetic field induced position fluctuations of a Sr MOT operated on the 689~nm transition. This achievement could improve for example the transfer efficiency of the ultracold atoms into far-off resonant dipole traps \cite{hasan2022anisotropic}. 

\textbf{Acknowledgments:} The authors thank Vladimir Akimov, David Petiteau, Riadh Rebhi, and Tridib Banerjee for their technical contributions to this work. This work was supported by the Centre for Quantum Technologies (CQT)/MoE funding Grant No. R-710-002-016-271, the Singapore Ministry of Education Academic Research Fund Tier2 Grant No. MOE2018-T2-1-082 (S), the National research foundation and Quantum Engineering Programme No.~NRF2021-QEP2-03-P01, and the French ANR Grant No. 16-CE30-0002-01. T.Z.-W. acknowledges Sorbonne Universit{\'e} and MajuLab for supporting a 12-month visiting research associate professorship at the Centre for Quantum Technologies (CQT) in Singapore.

\textbf{Conflict of interests:} The authors declare no conflict of interests

\textbf{Authors contribution:} JL, SD, CCK, AB, DW design and mount the experimental setup. JL, SD collect and analyse the data on the beam slower. KL, CCK collect and analyse the data on the magnetic field compensation. DW supervises the work. All authors contribute to the scientific discussion and to the writing of the manuscript.

\bibliography{SrII_setup}

\begin{thebibliography}{50}%
\makeatletter
\providecommand \@ifxundefined [1]{%
 \@ifx{#1\undefined}
}%
\providecommand \@ifnum [1]{%
 \ifnum #1\expandafter \@firstoftwo
 \else \expandafter \@secondoftwo
 \fi
}%
\providecommand \@ifx [1]{%
 \ifx #1\expandafter \@firstoftwo
 \else \expandafter \@secondoftwo
 \fi
}%
\providecommand \natexlab [1]{#1}%
\providecommand \enquote  [1]{``#1''}%
\providecommand \bibnamefont  [1]{#1}%
\providecommand \bibfnamefont [1]{#1}%
\providecommand \citenamefont [1]{#1}%
\providecommand \href@noop [0]{\@secondoftwo}%
\providecommand \href [0]{\begingroup \@sanitize@url \@href}%
\providecommand \@href[1]{\@@startlink{#1}\@@href}%
\providecommand \@@href[1]{\endgroup#1\@@endlink}%
\providecommand \@sanitize@url [0]{\catcode `\\12\catcode `\$12\catcode
  `\&12\catcode `\#12\catcode `\^12\catcode `\_12\catcode `\%12\relax}%
\providecommand \@@startlink[1]{}%
\providecommand \@@endlink[0]{}%
\providecommand \url  [0]{\begingroup\@sanitize@url \@url }%
\providecommand \@url [1]{\endgroup\@href {#1}{\urlprefix }}%
\providecommand \urlprefix  [0]{URL }%
\providecommand \Eprint [0]{\href }%
\providecommand \doibase [0]{https://doi.org/}%
\providecommand \selectlanguage [0]{\@gobble}%
\providecommand \bibinfo  [0]{\@secondoftwo}%
\providecommand \bibfield  [0]{\@secondoftwo}%
\providecommand \translation [1]{[#1]}%
\providecommand \BibitemOpen [0]{}%
\providecommand \bibitemStop [0]{}%
\providecommand \bibitemNoStop [0]{.\EOS\space}%
\providecommand \EOS [0]{\spacefactor3000\relax}%
\providecommand \BibitemShut  [1]{\csname bibitem#1\endcsname}%
\let\auto@bib@innerbib\@empty
\bibitem [{\citenamefont {Phillips}\ and\ \citenamefont
  {Metcalf}(1982)}]{Metcalf_lasercooling}%
  \BibitemOpen
  \bibfield  {author} {\bibinfo {author} {\bibfnamefont {W.~D.}\ \bibnamefont
  {Phillips}}\ and\ \bibinfo {author} {\bibfnamefont {H.}~\bibnamefont
  {Metcalf}},\ }\bibfield  {title} {\bibinfo {title} {Laser deceleration of an
  atomic beam},\ }\href {https://doi.org/10.1103/PhysRevLett.48.596} {\bibfield
   {journal} {\bibinfo  {journal} {Phys. Rev. Lett.}\ }\textbf {\bibinfo
  {volume} {48}},\ \bibinfo {pages} {596} (\bibinfo {year} {1982})}\BibitemShut
  {NoStop}%
\bibitem [{\citenamefont {Raab}\ \emph {et~al.}(1987)\citenamefont {Raab},
  \citenamefont {Prentiss}, \citenamefont {Cable}, \citenamefont {Chu},\ and\
  \citenamefont {Pritchard}}]{Pritchard_MOT}%
  \BibitemOpen
  \bibfield  {author} {\bibinfo {author} {\bibfnamefont {E.~L.}\ \bibnamefont
  {Raab}}, \bibinfo {author} {\bibfnamefont {M.}~\bibnamefont {Prentiss}},
  \bibinfo {author} {\bibfnamefont {A.}~\bibnamefont {Cable}}, \bibinfo
  {author} {\bibfnamefont {S.}~\bibnamefont {Chu}},\ and\ \bibinfo {author}
  {\bibfnamefont {D.~E.}\ \bibnamefont {Pritchard}},\ }\bibfield  {title}
  {\bibinfo {title} {Trapping of neutral sodium atoms with radiation
  pressure},\ }\href {https://doi.org/10.1103/PhysRevLett.59.2631} {\bibfield
  {journal} {\bibinfo  {journal} {Phys. Rev. Lett.}\ }\textbf {\bibinfo
  {volume} {59}},\ \bibinfo {pages} {2631} (\bibinfo {year}
  {1987})}\BibitemShut {NoStop}%
\bibitem [{\citenamefont {Guery-Odelin}\ and\ \citenamefont
  {Cohen-Tannoudji}(2011)}]{Guery-Odelin2011}%
  \BibitemOpen
  \bibfield  {author} {\bibinfo {author} {\bibfnamefont {D.}~\bibnamefont
  {Guery-Odelin}}\ and\ \bibinfo {author} {\bibfnamefont {C.}~\bibnamefont
  {Cohen-Tannoudji}},\ }\href@noop {} {\emph {\bibinfo {title} {Advances in
  atomic physics: an overview}}}\ (\bibinfo  {publisher} {World Scientific
  Publishing Co. Pte. Ltd.},\ \bibinfo {year} {2011})\BibitemShut {NoStop}%
\bibitem [{\citenamefont {Katori}\ \emph {et~al.}(2003)\citenamefont {Katori},
  \citenamefont {Takamoto}, \citenamefont {Pal'chikov},\ and\ \citenamefont
  {Ovsiannikov}}]{Katori2003}%
  \BibitemOpen
  \bibfield  {author} {\bibinfo {author} {\bibfnamefont {H.}~\bibnamefont
  {Katori}}, \bibinfo {author} {\bibfnamefont {M.}~\bibnamefont {Takamoto}},
  \bibinfo {author} {\bibfnamefont {V.}~\bibnamefont {Pal'chikov}},\ and\
  \bibinfo {author} {\bibfnamefont {V.}~\bibnamefont {Ovsiannikov}},\
  }\bibfield  {title} {\bibinfo {title} {Ultrastable optical clock with neutral
  atoms in an engineered light shift trap},\ }\href
  {https://doi.org/10.1103/PhysRevLett.91.173005} {\bibfield  {journal}
  {\bibinfo  {journal} {Phys. Rev. Lett.}\ }\textbf {\bibinfo {volume} {91}},\
  \bibinfo {pages} {173005} (\bibinfo {year} {2003})}\BibitemShut {NoStop}%
\bibitem [{\citenamefont {Ye}\ \emph {et~al.}(2008)\citenamefont {Ye},
  \citenamefont {Kimble},\ and\ \citenamefont {Katori}}]{Ye2008}%
  \BibitemOpen
  \bibfield  {author} {\bibinfo {author} {\bibfnamefont {J.}~\bibnamefont
  {Ye}}, \bibinfo {author} {\bibfnamefont {H.}~\bibnamefont {Kimble}},\ and\
  \bibinfo {author} {\bibfnamefont {H.}~\bibnamefont {Katori}},\ }\bibfield
  {title} {\bibinfo {title} {Quantum state engineering and precision metrology
  using state-insensitive light traps},\ }\href
  {https://doi.org/10.1126/science.1148259} {\bibfield  {journal} {\bibinfo
  {journal} {Science}\ }\textbf {\bibinfo {volume} {320}},\ \bibinfo {pages}
  {1734} (\bibinfo {year} {2008})}\BibitemShut {NoStop}%
\bibitem [{\citenamefont {Derevianko}\ and\ \citenamefont
  {Katori}(2011)}]{Derevianko2011}%
  \BibitemOpen
  \bibfield  {author} {\bibinfo {author} {\bibfnamefont {A.}~\bibnamefont
  {Derevianko}}\ and\ \bibinfo {author} {\bibfnamefont {H.}~\bibnamefont
  {Katori}},\ }\bibfield  {title} {\bibinfo {title} {Colloquium: Physics of
  optical lattice clocks},\ }\href {https://doi.org/10.1103/RevModPhys.83.331}
  {\bibfield  {journal} {\bibinfo  {journal} {Rev. Mod. Phys.}\ }\textbf
  {\bibinfo {volume} {83}},\ \bibinfo {pages} {331} (\bibinfo {year}
  {2011})}\BibitemShut {NoStop}%
\bibitem [{\citenamefont {Dalibard}\ \emph {et~al.}(2011)\citenamefont
  {Dalibard}, \citenamefont {Gerbier}, \citenamefont {Juzeli\~unas},\ and\
  \citenamefont {{\"O}hberg}}]{Dalibard2011}%
  \BibitemOpen
  \bibfield  {author} {\bibinfo {author} {\bibfnamefont {J.}~\bibnamefont
  {Dalibard}}, \bibinfo {author} {\bibfnamefont {F.}~\bibnamefont {Gerbier}},
  \bibinfo {author} {\bibfnamefont {G.}~\bibnamefont {Juzeli\~unas}},\ and\
  \bibinfo {author} {\bibfnamefont {P.}~\bibnamefont {{\"O}hberg}},\ }\bibfield
   {title} {\bibinfo {title} {Artificial gauge potentials for neutral atoms},\
  }\href {https://doi.org/10.1103/RevModPhys.83.1523} {\bibfield  {journal}
  {\bibinfo  {journal} {Rev. Mod. Phys.}\ }\textbf {\bibinfo {volume} {83}},\
  \bibinfo {pages} {1523} (\bibinfo {year} {2011})}\BibitemShut {NoStop}%
\bibitem [{\citenamefont {Bloch}\ \emph {et~al.}(2012)\citenamefont {Bloch},
  \citenamefont {Dalibard},\ and\ \citenamefont {Nascimb\`ene}}]{Bloch2012}%
  \BibitemOpen
  \bibfield  {author} {\bibinfo {author} {\bibfnamefont {I.}~\bibnamefont
  {Bloch}}, \bibinfo {author} {\bibfnamefont {J.}~\bibnamefont {Dalibard}},\
  and\ \bibinfo {author} {\bibfnamefont {S.}~\bibnamefont {Nascimb\`ene}},\
  }\bibfield  {title} {\bibinfo {title} {Quantum simulations with ultracold
  quantum gases},\ }\href {https://doi.org//10.1038/nphys2259} {\bibfield
  {journal} {\bibinfo  {journal} {Nat. Phys.}\ }\textbf {\bibinfo {volume}
  {8}},\ \bibinfo {pages} {267} (\bibinfo {year} {2012})}\BibitemShut {NoStop}%
\bibitem [{\citenamefont {Cooper}\ \emph {et~al.}(2019)\citenamefont {Cooper},
  \citenamefont {Dalibard},\ and\ \citenamefont {Spielman}}]{Cooper2019}%
  \BibitemOpen
  \bibfield  {author} {\bibinfo {author} {\bibfnamefont {N.}~\bibnamefont
  {Cooper}}, \bibinfo {author} {\bibfnamefont {J.}~\bibnamefont {Dalibard}},\
  and\ \bibinfo {author} {\bibfnamefont {I.}~\bibnamefont {Spielman}},\
  }\bibfield  {title} {\bibinfo {title} {Topological bands for ultracold
  atoms},\ }\href {https://doi.org/10.1103/RevModPhys.91.015005} {\bibfield
  {journal} {\bibinfo  {journal} {Rev. Mod. Phys.}\ }\textbf {\bibinfo {volume}
  {91}},\ \bibinfo {pages} {015005} (\bibinfo {year} {2019})}\BibitemShut
  {NoStop}%
\bibitem [{\citenamefont {Xu}\ \emph {et~al.}(2003)\citenamefont {Xu},
  \citenamefont {Loftus}, \citenamefont {Dunn}, \citenamefont {Greene},
  \citenamefont {Hall}, \citenamefont {Gallagher},\ and\ \citenamefont
  {Ye}}]{Xu2003}%
  \BibitemOpen
  \bibfield  {author} {\bibinfo {author} {\bibfnamefont {X.}~\bibnamefont
  {Xu}}, \bibinfo {author} {\bibfnamefont {T.}~\bibnamefont {Loftus}}, \bibinfo
  {author} {\bibfnamefont {J.}~\bibnamefont {Dunn}}, \bibinfo {author}
  {\bibfnamefont {C.}~\bibnamefont {Greene}}, \bibinfo {author} {\bibfnamefont
  {J.}~\bibnamefont {Hall}}, \bibinfo {author} {\bibfnamefont {A.}~\bibnamefont
  {Gallagher}},\ and\ \bibinfo {author} {\bibfnamefont {J.}~\bibnamefont
  {Ye}},\ }\bibfield  {title} {\bibinfo {title} {Single-stage sub-doppler
  cooling of alkaline earth atoms},\ }\href
  {https://doi.org/10.1103/PhysRevLett.90.193002} {\bibfield  {journal}
  {\bibinfo  {journal} {Phys. Rev. Lett.}\ }\textbf {\bibinfo {volume} {90}},\
  \bibinfo {pages} {193002} (\bibinfo {year} {2003})}\BibitemShut {NoStop}%
\bibitem [{\citenamefont {Loftus}\ \emph {et~al.}(2004)\citenamefont {Loftus},
  \citenamefont {Ido}, \citenamefont {Boyd}, \citenamefont {Ludlow},\ and\
  \citenamefont {Ye}}]{Loftus2004}%
  \BibitemOpen
  \bibfield  {author} {\bibinfo {author} {\bibfnamefont {T.}~\bibnamefont
  {Loftus}}, \bibinfo {author} {\bibfnamefont {T.}~\bibnamefont {Ido}},
  \bibinfo {author} {\bibfnamefont {M.}~\bibnamefont {Boyd}}, \bibinfo {author}
  {\bibfnamefont {A.}~\bibnamefont {Ludlow}},\ and\ \bibinfo {author}
  {\bibfnamefont {J.}~\bibnamefont {Ye}},\ }\bibfield  {title} {\bibinfo
  {title} {Narrow line cooling and momentum-space crystals},\ }\href
  {https://doi.org/10.1103/PhysRevA.70.063413} {\bibfield  {journal} {\bibinfo
  {journal} {Phys. Rev. A}\ }\textbf {\bibinfo {volume} {70}},\ \bibinfo
  {pages} {063413} (\bibinfo {year} {2004})}\BibitemShut {NoStop}%
\bibitem [{\citenamefont {Chaneliere}\ \emph {et~al.}(2008)\citenamefont
  {Chaneliere}, \citenamefont {He}, \citenamefont {Kaiser},\ and\ \citenamefont
  {Wilkowski}}]{chaneliere2008three}%
  \BibitemOpen
  \bibfield  {author} {\bibinfo {author} {\bibfnamefont {T.}~\bibnamefont
  {Chaneliere}}, \bibinfo {author} {\bibfnamefont {L.}~\bibnamefont {He}},
  \bibinfo {author} {\bibfnamefont {R.}~\bibnamefont {Kaiser}},\ and\ \bibinfo
  {author} {\bibfnamefont {D.}~\bibnamefont {Wilkowski}},\ }\bibfield  {title}
  {\bibinfo {title} {Three dimensional cooling and trapping with a narrow
  line},\ }\href {https://doi.org/https://doi.org/10.1140/epjd/e2007-00329-8}
  {\bibfield  {journal} {\bibinfo  {journal} {Eur. Phys. J. D}\ }\textbf
  {\bibinfo {volume} {46}},\ \bibinfo {pages} {507} (\bibinfo {year}
  {2008})}\BibitemShut {NoStop}%
\bibitem [{\citenamefont {Chalony}\ \emph {et~al.}(2011)\citenamefont
  {Chalony}, \citenamefont {Kastberg}, \citenamefont {Klappauf},\ and\
  \citenamefont {Wilkowski}}]{chalony2011doppler}%
  \BibitemOpen
  \bibfield  {author} {\bibinfo {author} {\bibfnamefont {M.}~\bibnamefont
  {Chalony}}, \bibinfo {author} {\bibfnamefont {A.}~\bibnamefont {Kastberg}},
  \bibinfo {author} {\bibfnamefont {B.}~\bibnamefont {Klappauf}},\ and\
  \bibinfo {author} {\bibfnamefont {D.}~\bibnamefont {Wilkowski}},\ }\bibfield
  {title} {\bibinfo {title} {Doppler cooling to the quantum limit},\ }\href
  {https://doi.org/10.1103/PhysRevLett.107.243002} {\bibfield  {journal}
  {\bibinfo  {journal} {Phys. Rev. Lett.}\ }\textbf {\bibinfo {volume} {107}},\
  \bibinfo {pages} {243002} (\bibinfo {year} {2011})}\BibitemShut {NoStop}%
\bibitem [{\citenamefont {Boyd}\ \emph {et~al.}(2007)\citenamefont {Boyd},
  \citenamefont {Zelevinsky}, \citenamefont {Ludlow}, \citenamefont {Blatt},
  \citenamefont {Zanon-Willette}, \citenamefont {Foreman},\ and\ \citenamefont
  {Ye}}]{Marty2007}%
  \BibitemOpen
  \bibfield  {author} {\bibinfo {author} {\bibfnamefont {M.~M.}\ \bibnamefont
  {Boyd}}, \bibinfo {author} {\bibfnamefont {T.}~\bibnamefont {Zelevinsky}},
  \bibinfo {author} {\bibfnamefont {A.~D.}\ \bibnamefont {Ludlow}}, \bibinfo
  {author} {\bibfnamefont {S.}~\bibnamefont {Blatt}}, \bibinfo {author}
  {\bibfnamefont {T.}~\bibnamefont {Zanon-Willette}}, \bibinfo {author}
  {\bibfnamefont {S.~M.}\ \bibnamefont {Foreman}},\ and\ \bibinfo {author}
  {\bibfnamefont {J.}~\bibnamefont {Ye}},\ }\bibfield  {title} {\bibinfo
  {title} {Nuclear spin effects in optical lattice clocks},\ }\href
  {https://doi.org/10.1103/PhysRevA.76.022510} {\bibfield  {journal} {\bibinfo
  {journal} {Phys. Rev. A}\ }\textbf {\bibinfo {volume} {76}},\ \bibinfo
  {pages} {022510} (\bibinfo {year} {2007})}\BibitemShut {NoStop}%
\bibitem [{\citenamefont {Taichenachev}\ \emph {et~al.}(2006)\citenamefont
  {Taichenachev}, \citenamefont {Yudin}, \citenamefont {Oates}, \citenamefont
  {Hoyt}, \citenamefont {Barber},\ and\ \citenamefont
  {Hollberg}}]{Taichenachev2006}%
  \BibitemOpen
  \bibfield  {author} {\bibinfo {author} {\bibfnamefont {A.}~\bibnamefont
  {Taichenachev}}, \bibinfo {author} {\bibfnamefont {V.}~\bibnamefont {Yudin}},
  \bibinfo {author} {\bibfnamefont {C.}~\bibnamefont {Oates}}, \bibinfo
  {author} {\bibfnamefont {C.}~\bibnamefont {Hoyt}}, \bibinfo {author}
  {\bibfnamefont {Z.}~\bibnamefont {Barber}},\ and\ \bibinfo {author}
  {\bibfnamefont {L.}~\bibnamefont {Hollberg}},\ }\bibfield  {title} {\bibinfo
  {title} {Magnetic field-induced spectroscopy of forbidden optical transitions
  with application to lattice-based optical atomic clocks},\ }\href
  {https://doi.org/10.1103/PhysRevLett.96.083001} {\bibfield  {journal}
  {\bibinfo  {journal} {Phys. Rev. Lett.}\ }\textbf {\bibinfo {volume} {96}},\
  \bibinfo {pages} {083001} (\bibinfo {year} {2006})}\BibitemShut {NoStop}%
\bibitem [{\citenamefont {Santra}\ \emph {et~al.}(2005)\citenamefont {Santra},
  \citenamefont {Arimondo}, \citenamefont {Ido}, \citenamefont {Greene},\ and\
  \citenamefont {Ye}}]{Santra2005}%
  \BibitemOpen
  \bibfield  {author} {\bibinfo {author} {\bibfnamefont {R.}~\bibnamefont
  {Santra}}, \bibinfo {author} {\bibfnamefont {E.}~\bibnamefont {Arimondo}},
  \bibinfo {author} {\bibfnamefont {T.}~\bibnamefont {Ido}}, \bibinfo {author}
  {\bibfnamefont {C.}~\bibnamefont {Greene}},\ and\ \bibinfo {author}
  {\bibfnamefont {J.}~\bibnamefont {Ye}},\ }\bibfield  {title} {\bibinfo
  {title} {High-accuracy optical clock via three-level coherence in neutral
  bosonic $^{88}${S}r},\ }\href {https://doi.org/10.1103/PhysRevLett.94.173002}
  {\bibfield  {journal} {\bibinfo  {journal} {Phys. Rev. Lett.}\ }\textbf
  {\bibinfo {volume} {94}},\ \bibinfo {pages} {173002} (\bibinfo {year}
  {2005})}\BibitemShut {NoStop}%
\bibitem [{\citenamefont {Hong}\ \emph {et~al.}(2005)\citenamefont {Hong},
  \citenamefont {Cramer}, \citenamefont {Nagourney},\ and\ \citenamefont
  {Fortson}}]{Hong2005}%
  \BibitemOpen
  \bibfield  {author} {\bibinfo {author} {\bibfnamefont {T.}~\bibnamefont
  {Hong}}, \bibinfo {author} {\bibfnamefont {C.}~\bibnamefont {Cramer}},
  \bibinfo {author} {\bibfnamefont {W.}~\bibnamefont {Nagourney}},\ and\
  \bibinfo {author} {\bibfnamefont {E.}~\bibnamefont {Fortson}},\ }\bibfield
  {title} {\bibinfo {title} {Optical clocks based on ultranarrow three-photon
  resonances in alkaline earth atoms},\ }\href
  {https://doi.org/10.1103/PhysRevLett.94.050801} {\bibfield  {journal}
  {\bibinfo  {journal} {Phys. Rev. Lett.}\ }\textbf {\bibinfo {volume} {94}},\
  \bibinfo {pages} {050801} (\bibinfo {year} {2005})}\BibitemShut {NoStop}%
\bibitem [{\citenamefont {Baillard}\ \emph {et~al.}(2007)\citenamefont
  {Baillard}, \citenamefont {Fouch\'{e}}, \citenamefont {Targat}, \citenamefont
  {Westergaard}, \citenamefont {Lecallier}, \citenamefont {Coq}, \citenamefont
  {Rovera}, \citenamefont {Bize},\ and\ \citenamefont {Lemonde}}]{magic_B}%
  \BibitemOpen
  \bibfield  {author} {\bibinfo {author} {\bibfnamefont {X.}~\bibnamefont
  {Baillard}}, \bibinfo {author} {\bibfnamefont {M.}~\bibnamefont
  {Fouch\'{e}}}, \bibinfo {author} {\bibfnamefont {R.~L.}\ \bibnamefont
  {Targat}}, \bibinfo {author} {\bibfnamefont {P.~G.}\ \bibnamefont
  {Westergaard}}, \bibinfo {author} {\bibfnamefont {A.}~\bibnamefont
  {Lecallier}}, \bibinfo {author} {\bibfnamefont {Y.~L.}\ \bibnamefont {Coq}},
  \bibinfo {author} {\bibfnamefont {G.~D.}\ \bibnamefont {Rovera}}, \bibinfo
  {author} {\bibfnamefont {S.}~\bibnamefont {Bize}},\ and\ \bibinfo {author}
  {\bibfnamefont {P.}~\bibnamefont {Lemonde}},\ }\bibfield  {title} {\bibinfo
  {title} {Accuracy evaluation of an optical lattice clock with bosonic
  atoms},\ }\href {https://doi.org/10.1364/OL.32.001812} {\bibfield  {journal}
  {\bibinfo  {journal} {Opt. Lett.}\ }\textbf {\bibinfo {volume} {32}},\
  \bibinfo {pages} {1812} (\bibinfo {year} {2007})}\BibitemShut {NoStop}%
\bibitem [{\citenamefont {Ovsiannikov}\ \emph {et~al.}(2007)\citenamefont
  {Ovsiannikov}, \citenamefont {Pal'chikov}, \citenamefont {Taichenachev},
  \citenamefont {Yudin}, \citenamefont {Katori},\ and\ \citenamefont
  {Takamoto}}]{Katori_magic_circular}%
  \BibitemOpen
  \bibfield  {author} {\bibinfo {author} {\bibfnamefont {V.~D.}\ \bibnamefont
  {Ovsiannikov}}, \bibinfo {author} {\bibfnamefont {V.~G.}\ \bibnamefont
  {Pal'chikov}}, \bibinfo {author} {\bibfnamefont {A.~V.}\ \bibnamefont
  {Taichenachev}}, \bibinfo {author} {\bibfnamefont {V.~I.}\ \bibnamefont
  {Yudin}}, \bibinfo {author} {\bibfnamefont {H.}~\bibnamefont {Katori}},\ and\
  \bibinfo {author} {\bibfnamefont {M.}~\bibnamefont {Takamoto}},\ }\bibfield
  {title} {\bibinfo {title} {Magic-wave-induced
  $^{1}\mathbf{S}_{0}\text{\ensuremath{-}}^{3}\mathbf{P}_{0}$ transition in
  even isotopes of alkaline-earth-metal-like atoms},\ }\href
  {https://doi.org/10.1103/PhysRevA.75.020501} {\bibfield  {journal} {\bibinfo
  {journal} {Phys. Rev. A}\ }\textbf {\bibinfo {volume} {75}},\ \bibinfo
  {pages} {020501} (\bibinfo {year} {2007})}\BibitemShut {NoStop}%
\bibitem [{\citenamefont {Akatsuka}\ \emph {et~al.}(2008)\citenamefont
  {Akatsuka}, \citenamefont {Takamoto},\ and\ \citenamefont
  {Katori}}]{Akatsuka2008}%
  \BibitemOpen
  \bibfield  {author} {\bibinfo {author} {\bibfnamefont {T.}~\bibnamefont
  {Akatsuka}}, \bibinfo {author} {\bibfnamefont {M.}~\bibnamefont {Takamoto}},\
  and\ \bibinfo {author} {\bibfnamefont {H.}~\bibnamefont {Katori}},\
  }\bibfield  {title} {\bibinfo {title} {Optical lattice clocks with
  non-interacting bosons and fermions},\ }\href
  {https://doi.org/10.1038/nphys1108} {\bibfield  {journal} {\bibinfo
  {journal} {Nat. Phys.}\ }\textbf {\bibinfo {volume} {4}},\ \bibinfo {pages}
  {954} (\bibinfo {year} {2008})}\BibitemShut {NoStop}%
\bibitem [{\citenamefont {Le~Targat}\ \emph {et~al.}(2013)\citenamefont
  {Le~Targat}, \citenamefont {Lorini}, \citenamefont {Le~Coq}, \citenamefont
  {Zawada}, \citenamefont {Gu\'ena}, \citenamefont {Abgrall}, \citenamefont
  {Gurov}, \citenamefont {Rosenbusch}, \citenamefont {Rovera}, \citenamefont
  {Nag\'orny}, \citenamefont {Gartman}, \citenamefont {Westergaard},
  \citenamefont {Tobar}, \citenamefont {Lours}, \citenamefont {Santarelli},
  \citenamefont {Clairon}, \citenamefont {Bize}, \citenamefont {Laurent},
  \citenamefont {Lemonde},\ and\ \citenamefont {Lodewyck}}]{Letargat2013}%
  \BibitemOpen
  \bibfield  {author} {\bibinfo {author} {\bibfnamefont {R.}~\bibnamefont
  {Le~Targat}}, \bibinfo {author} {\bibfnamefont {L.}~\bibnamefont {Lorini}},
  \bibinfo {author} {\bibfnamefont {Y.}~\bibnamefont {Le~Coq}}, \bibinfo
  {author} {\bibfnamefont {M.}~\bibnamefont {Zawada}}, \bibinfo {author}
  {\bibfnamefont {J.}~\bibnamefont {Gu\'ena}}, \bibinfo {author} {\bibfnamefont
  {M.}~\bibnamefont {Abgrall}}, \bibinfo {author} {\bibfnamefont
  {M.}~\bibnamefont {Gurov}}, \bibinfo {author} {\bibfnamefont
  {P.}~\bibnamefont {Rosenbusch}}, \bibinfo {author} {\bibfnamefont
  {D.}~\bibnamefont {Rovera}}, \bibinfo {author} {\bibfnamefont
  {B.}~\bibnamefont {Nag\'orny}}, \bibinfo {author} {\bibfnamefont
  {R.}~\bibnamefont {Gartman}}, \bibinfo {author} {\bibfnamefont
  {P.}~\bibnamefont {Westergaard}}, \bibinfo {author} {\bibfnamefont
  {M.}~\bibnamefont {Tobar}}, \bibinfo {author} {\bibfnamefont
  {M.}~\bibnamefont {Lours}}, \bibinfo {author} {\bibfnamefont
  {G.}~\bibnamefont {Santarelli}}, \bibinfo {author} {\bibfnamefont
  {A.}~\bibnamefont {Clairon}}, \bibinfo {author} {\bibfnamefont
  {S.}~\bibnamefont {Bize}}, \bibinfo {author} {\bibfnamefont {P.}~\bibnamefont
  {Laurent}}, \bibinfo {author} {\bibfnamefont {P.}~\bibnamefont {Lemonde}},\
  and\ \bibinfo {author} {\bibfnamefont {J.}~\bibnamefont {Lodewyck}},\
  }\bibfield  {title} {\bibinfo {title} {Experimental realization of an optical
  second with strontium lattice clocks},\ }\href
  {https://doi.org/10.1038/ncomms3109} {\bibfield  {journal} {\bibinfo
  {journal} {Nat. Commun.}\ }\textbf {\bibinfo {volume} {4}},\ \bibinfo {pages}
  {2109} (\bibinfo {year} {2013})}\BibitemShut {NoStop}%
\bibitem [{\citenamefont {Ludlow}\ \emph {et~al.}(2015)\citenamefont {Ludlow},
  \citenamefont {Boyd}, \citenamefont {Ye}, \citenamefont {Peik},\ and\
  \citenamefont {Schmidt}}]{Ludlow_clock}%
  \BibitemOpen
  \bibfield  {author} {\bibinfo {author} {\bibfnamefont {A.~D.}\ \bibnamefont
  {Ludlow}}, \bibinfo {author} {\bibfnamefont {M.~M.}\ \bibnamefont {Boyd}},
  \bibinfo {author} {\bibfnamefont {J.}~\bibnamefont {Ye}}, \bibinfo {author}
  {\bibfnamefont {E.}~\bibnamefont {Peik}},\ and\ \bibinfo {author}
  {\bibfnamefont {P.~O.}\ \bibnamefont {Schmidt}},\ }\bibfield  {title}
  {\bibinfo {title} {Optical atomic clocks},\ }\href
  {https://doi.org/10.1103/RevModPhys.87.637} {\bibfield  {journal} {\bibinfo
  {journal} {Rev. Mod. Phys.}\ }\textbf {\bibinfo {volume} {87}},\ \bibinfo
  {pages} {637} (\bibinfo {year} {2015})}\BibitemShut {NoStop}%
\bibitem [{\citenamefont {Bothwell}\ \emph {et~al.}(2019)\citenamefont
  {Bothwell}, \citenamefont {Kedar}, \citenamefont {Oelker}, \citenamefont
  {Robinson}, \citenamefont {Bromley}, \citenamefont {Tew}, \citenamefont
  {Ye},\ and\ \citenamefont {Kennedy}}]{Bothwell2019}%
  \BibitemOpen
  \bibfield  {author} {\bibinfo {author} {\bibfnamefont {T.}~\bibnamefont
  {Bothwell}}, \bibinfo {author} {\bibfnamefont {D.}~\bibnamefont {Kedar}},
  \bibinfo {author} {\bibfnamefont {E.}~\bibnamefont {Oelker}}, \bibinfo
  {author} {\bibfnamefont {J.~M.}\ \bibnamefont {Robinson}}, \bibinfo {author}
  {\bibfnamefont {S.~L.}\ \bibnamefont {Bromley}}, \bibinfo {author}
  {\bibfnamefont {W.~L.}\ \bibnamefont {Tew}}, \bibinfo {author} {\bibfnamefont
  {J.}~\bibnamefont {Ye}},\ and\ \bibinfo {author} {\bibfnamefont {C.~J.}\
  \bibnamefont {Kennedy}},\ }\bibfield  {title} {\bibinfo {title} {\textup{JILA
  SrI} optical lattice clock with uncertainty of $2.0\times10^{-18}$},\ }\href
  {https://doi.org/10.1088/1681-7575/ab4089} {\bibfield  {journal} {\bibinfo
  {journal} {Metrologia}\ }\textbf {\bibinfo {volume} {56}},\ \bibinfo {pages}
  {065004} (\bibinfo {year} {2019})}\BibitemShut {NoStop}%
\bibitem [{\citenamefont {Hu}\ \emph {et~al.}(2017)\citenamefont {Hu},
  \citenamefont {Poli}, \citenamefont {Salvi},\ and\ \citenamefont
  {Tino}}]{Tino_interferometer}%
  \BibitemOpen
  \bibfield  {author} {\bibinfo {author} {\bibfnamefont {L.}~\bibnamefont
  {Hu}}, \bibinfo {author} {\bibfnamefont {N.}~\bibnamefont {Poli}}, \bibinfo
  {author} {\bibfnamefont {L.}~\bibnamefont {Salvi}},\ and\ \bibinfo {author}
  {\bibfnamefont {G.~M.}\ \bibnamefont {Tino}},\ }\bibfield  {title} {\bibinfo
  {title} {Atom interferometry with the sr optical clock transition},\ }\href
  {https://doi.org/10.1103/PhysRevLett.119.263601} {\bibfield  {journal}
  {\bibinfo  {journal} {Phys. Rev. Lett.}\ }\textbf {\bibinfo {volume} {119}},\
  \bibinfo {pages} {263601} (\bibinfo {year} {2017})}\BibitemShut {NoStop}%
\bibitem [{\citenamefont {Graham}\ \emph {et~al.}(2013)\citenamefont {Graham},
  \citenamefont {Hogan}, \citenamefont {Kasevich},\ and\ \citenamefont
  {Rajendran}}]{graham2013}%
  \BibitemOpen
  \bibfield  {author} {\bibinfo {author} {\bibfnamefont {P.~W.}\ \bibnamefont
  {Graham}}, \bibinfo {author} {\bibfnamefont {J.~M.}\ \bibnamefont {Hogan}},
  \bibinfo {author} {\bibfnamefont {M.~A.}\ \bibnamefont {Kasevich}},\ and\
  \bibinfo {author} {\bibfnamefont {S.}~\bibnamefont {Rajendran}},\ }\bibfield
  {title} {\bibinfo {title} {New method for gravitational wave detection with
  atomic sensors},\ }\href {https://doi.org/10.1103/PhysRevLett.110.171102}
  {\bibfield  {journal} {\bibinfo  {journal} {Phys. Rev. Lett.}\ }\textbf
  {\bibinfo {volume} {110}},\ \bibinfo {pages} {171102} (\bibinfo {year}
  {2013})}\BibitemShut {NoStop}%
\bibitem [{\citenamefont {Kolkowitz}\ \emph {et~al.}(2017)\citenamefont
  {Kolkowitz}, \citenamefont {Bromley}, \citenamefont {Wall}, \citenamefont
  {Marti}, \citenamefont {Koller}, \citenamefont {Zhang}, \citenamefont {Rey},\
  and\ \citenamefont {Ye}}]{Kolkowitz_Qsimulation}%
  \BibitemOpen
  \bibfield  {author} {\bibinfo {author} {\bibfnamefont {S.}~\bibnamefont
  {Kolkowitz}}, \bibinfo {author} {\bibfnamefont {T.}~\bibnamefont {Bromley},
  \bibfnamefont {S.~L.and~Bothwell}}, \bibinfo {author} {\bibfnamefont {M.~L.}\
  \bibnamefont {Wall}}, \bibinfo {author} {\bibfnamefont {G.~E.}\ \bibnamefont
  {Marti}}, \bibinfo {author} {\bibfnamefont {A.~P.}\ \bibnamefont {Koller}},
  \bibinfo {author} {\bibfnamefont {X.}~\bibnamefont {Zhang}}, \bibinfo
  {author} {\bibfnamefont {A.~M.}\ \bibnamefont {Rey}},\ and\ \bibinfo {author}
  {\bibfnamefont {J.}~\bibnamefont {Ye}},\ }\bibfield  {title} {\bibinfo
  {title} {Spin-orbit-coupled fermions in an optical lattice clock},\ }\href
  {https://doi.org/10.1038/nature20811} {\bibfield  {journal} {\bibinfo
  {journal} {Nature (London)}\ }\textbf {\bibinfo {volume} {542}},\ \bibinfo
  {pages} {66} (\bibinfo {year} {2017})}\BibitemShut {NoStop}%
\bibitem [{\citenamefont {Rajagopal}\ \emph {et~al.}(2017)\citenamefont
  {Rajagopal}, \citenamefont {Fujiwara}, \citenamefont {Senaratne},
  \citenamefont {Singh}, \citenamefont {Geiger},\ and\ \citenamefont
  {Weld}}]{Rajagopal_Qsimulation}%
  \BibitemOpen
  \bibfield  {author} {\bibinfo {author} {\bibfnamefont {S.~V.}\ \bibnamefont
  {Rajagopal}}, \bibinfo {author} {\bibfnamefont {K.~M.}\ \bibnamefont
  {Fujiwara}}, \bibinfo {author} {\bibfnamefont {R.}~\bibnamefont {Senaratne}},
  \bibinfo {author} {\bibfnamefont {K.}~\bibnamefont {Singh}}, \bibinfo
  {author} {\bibfnamefont {Z.~A.}\ \bibnamefont {Geiger}},\ and\ \bibinfo
  {author} {\bibfnamefont {D.~M.}\ \bibnamefont {Weld}},\ }\bibfield  {title}
  {\bibinfo {title} {Quantum emulation of extreme non-equilibrium phenomena
  with trapped atoms},\ }\href {https://doi.org/10.1002/andp.201700008}
  {\bibfield  {journal} {\bibinfo  {journal} {Ann, Phys.}\ }\textbf {\bibinfo
  {volume} {529}},\ \bibinfo {pages} {1700008} (\bibinfo {year}
  {2017})}\BibitemShut {NoStop}%
\bibitem [{\citenamefont {Heinz}\ \emph {et~al.}(2020)\citenamefont {Heinz},
  \citenamefont {Park}, \citenamefont {\ifmmode \check{S}\else
  \v{S}\fi{}anti\ifmmode~\acute{c}\else \'{c}\fi{}}, \citenamefont {Trautmann},
  \citenamefont {Porsev}, \citenamefont {Safronova}, \citenamefont {Bloch},\
  and\ \citenamefont {Blatt}}]{Blatt_Qcomputation}%
  \BibitemOpen
  \bibfield  {author} {\bibinfo {author} {\bibfnamefont {A.}~\bibnamefont
  {Heinz}}, \bibinfo {author} {\bibfnamefont {A.~J.}\ \bibnamefont {Park}},
  \bibinfo {author} {\bibfnamefont {N.}~\bibnamefont {\ifmmode \check{S}\else
  \v{S}\fi{}anti\ifmmode~\acute{c}\else \'{c}\fi{}}}, \bibinfo {author}
  {\bibfnamefont {J.}~\bibnamefont {Trautmann}}, \bibinfo {author}
  {\bibfnamefont {S.~G.}\ \bibnamefont {Porsev}}, \bibinfo {author}
  {\bibfnamefont {M.~S.}\ \bibnamefont {Safronova}}, \bibinfo {author}
  {\bibfnamefont {I.}~\bibnamefont {Bloch}},\ and\ \bibinfo {author}
  {\bibfnamefont {S.}~\bibnamefont {Blatt}},\ }\bibfield  {title} {\bibinfo
  {title} {State-dependent optical lattices for the strontium optical qubit},\
  }\href {https://doi.org/10.1103/PhysRevLett.124.203201} {\bibfield  {journal}
  {\bibinfo  {journal} {Phys. Rev. Lett.}\ }\textbf {\bibinfo {volume} {124}},\
  \bibinfo {pages} {203201} (\bibinfo {year} {2020})}\BibitemShut {NoStop}%
\bibitem [{\citenamefont {Daley}\ \emph {et~al.}(2008)\citenamefont {Daley},
  \citenamefont {Boyd}, \citenamefont {Ye},\ and\ \citenamefont
  {Zoller}}]{PeterZoller_Qcomputation}%
  \BibitemOpen
  \bibfield  {author} {\bibinfo {author} {\bibfnamefont {A.~J.}\ \bibnamefont
  {Daley}}, \bibinfo {author} {\bibfnamefont {M.~M.}\ \bibnamefont {Boyd}},
  \bibinfo {author} {\bibfnamefont {J.}~\bibnamefont {Ye}},\ and\ \bibinfo
  {author} {\bibfnamefont {P.}~\bibnamefont {Zoller}},\ }\bibfield  {title}
  {\bibinfo {title} {Quantum computing with alkaline-earth-metal atoms},\
  }\href {https://doi.org/10.1103/PhysRevLett.101.170504} {\bibfield  {journal}
  {\bibinfo  {journal} {Phys. Rev. Lett.}\ }\textbf {\bibinfo {volume} {101}},\
  \bibinfo {pages} {170504} (\bibinfo {year} {2008})}\BibitemShut {NoStop}%
\bibitem [{\citenamefont {Gorshkov}\ \emph {et~al.}(2009)\citenamefont
  {Gorshkov}, \citenamefont {Rey}, \citenamefont {Daley}, \citenamefont {Boyd},
  \citenamefont {Ye}, \citenamefont {Zoller},\ and\ \citenamefont
  {Lukin}}]{Lukin_Qcomputatiom}%
  \BibitemOpen
  \bibfield  {author} {\bibinfo {author} {\bibfnamefont {A.~V.}\ \bibnamefont
  {Gorshkov}}, \bibinfo {author} {\bibfnamefont {A.~M.}\ \bibnamefont {Rey}},
  \bibinfo {author} {\bibfnamefont {A.~J.}\ \bibnamefont {Daley}}, \bibinfo
  {author} {\bibfnamefont {M.~M.}\ \bibnamefont {Boyd}}, \bibinfo {author}
  {\bibfnamefont {J.}~\bibnamefont {Ye}}, \bibinfo {author} {\bibfnamefont
  {P.}~\bibnamefont {Zoller}},\ and\ \bibinfo {author} {\bibfnamefont {M.~D.}\
  \bibnamefont {Lukin}},\ }\bibfield  {title} {\bibinfo {title}
  {Alkaline-earth-metal atoms as few-qubit quantum registers},\ }\href
  {https://doi.org/10.1103/PhysRevLett.102.11050} {\bibfield  {journal}
  {\bibinfo  {journal} {Phys. Rev. Lett.}\ }\textbf {\bibinfo {volume} {102}},\
  \bibinfo {pages} {110503} (\bibinfo {year} {2009})}\BibitemShut {NoStop}%
\bibitem [{\citenamefont {Yang}\ \emph {et~al.}(2015)\citenamefont {Yang},
  \citenamefont {Pandey}, \citenamefont {Pramod}, \citenamefont {Leroux},
  \citenamefont {Kwong}, \citenamefont {Hajiyev}, \citenamefont {Chia},
  \citenamefont {Fang},\ and\ \citenamefont {Wilkowski}}]{yang2015high}%
  \BibitemOpen
  \bibfield  {author} {\bibinfo {author} {\bibfnamefont {T.}~\bibnamefont
  {Yang}}, \bibinfo {author} {\bibfnamefont {K.}~\bibnamefont {Pandey}},
  \bibinfo {author} {\bibfnamefont {M.~S.}\ \bibnamefont {Pramod}}, \bibinfo
  {author} {\bibfnamefont {F.}~\bibnamefont {Leroux}}, \bibinfo {author}
  {\bibfnamefont {C.~C.}\ \bibnamefont {Kwong}}, \bibinfo {author}
  {\bibfnamefont {E.}~\bibnamefont {Hajiyev}}, \bibinfo {author} {\bibfnamefont
  {Z.~Y.}\ \bibnamefont {Chia}}, \bibinfo {author} {\bibfnamefont
  {B.}~\bibnamefont {Fang}},\ and\ \bibinfo {author} {\bibfnamefont
  {D.}~\bibnamefont {Wilkowski}},\ }\bibfield  {title} {\bibinfo {title} {A
  high flux source of cold strontium atoms},\ }\href
  {https://doi.org/10.1140/epjd/e2015-60288-y} {\bibfield  {journal} {\bibinfo
  {journal} {Eur. Phys. J. D}\ }\textbf {\bibinfo {volume} {69}},\ \bibinfo
  {pages} {226} (\bibinfo {year} {2015})}\BibitemShut {NoStop}%
\bibitem [{\citenamefont {Tiecke}\ \emph {et~al.}(2009)\citenamefont {Tiecke},
  \citenamefont {Gensemer}, \citenamefont {Ludewig},\ and\ \citenamefont
  {Walraven}}]{Walraven2009Li}%
  \BibitemOpen
  \bibfield  {author} {\bibinfo {author} {\bibfnamefont {T.~G.}\ \bibnamefont
  {Tiecke}}, \bibinfo {author} {\bibfnamefont {S.~D.}\ \bibnamefont
  {Gensemer}}, \bibinfo {author} {\bibfnamefont {A.}~\bibnamefont {Ludewig}},\
  and\ \bibinfo {author} {\bibfnamefont {J.~T.~M.}\ \bibnamefont {Walraven}},\
  }\bibfield  {title} {\bibinfo {title} {High-flux two-dimensional
  magneto-optical-trap source for cold lithium atoms},\ }\href
  {https://doi.org/10.1103/PhysRevA.80.013409} {\bibfield  {journal} {\bibinfo
  {journal} {Phys. Rev. A}\ }\textbf {\bibinfo {volume} {80}},\ \bibinfo
  {pages} {013409} (\bibinfo {year} {2009})}\BibitemShut {NoStop}%
\bibitem [{\citenamefont {Lamporesi}\ \emph {et~al.}(2013)\citenamefont
  {Lamporesi}, \citenamefont {Donadello}, \citenamefont {Serafini},\ and\
  \citenamefont {Ferrari}}]{lamporesi2013compact}%
  \BibitemOpen
  \bibfield  {author} {\bibinfo {author} {\bibfnamefont {G.}~\bibnamefont
  {Lamporesi}}, \bibinfo {author} {\bibfnamefont {S.}~\bibnamefont
  {Donadello}}, \bibinfo {author} {\bibfnamefont {S.}~\bibnamefont
  {Serafini}},\ and\ \bibinfo {author} {\bibfnamefont {G.}~\bibnamefont
  {Ferrari}},\ }\bibfield  {title} {\bibinfo {title} {Compact high-flux source
  of cold sodium atoms},\ }\href {https://doi.org/10.1063/1.4808375} {\bibfield
   {journal} {\bibinfo  {journal} {Rev. Sci. Instrum.}\ }\textbf {\bibinfo
  {volume} {84}},\ \bibinfo {pages} {063102} (\bibinfo {year}
  {2013})}\BibitemShut {NoStop}%
\bibitem [{\citenamefont {Nosske}\ \emph {et~al.}(2017)\citenamefont {Nosske},
  \citenamefont {Couturier}, \citenamefont {Hu}, \citenamefont {Tan},
  \citenamefont {Qiao}, \citenamefont {Blume}, \citenamefont {Jiang},
  \citenamefont {Chen},\ and\ \citenamefont {Weidem\"uller}}]{Widemuller_Sr}%
  \BibitemOpen
  \bibfield  {author} {\bibinfo {author} {\bibfnamefont {I.}~\bibnamefont
  {Nosske}}, \bibinfo {author} {\bibfnamefont {L.}~\bibnamefont {Couturier}},
  \bibinfo {author} {\bibfnamefont {F.}~\bibnamefont {Hu}}, \bibinfo {author}
  {\bibfnamefont {C.}~\bibnamefont {Tan}}, \bibinfo {author} {\bibfnamefont
  {C.}~\bibnamefont {Qiao}}, \bibinfo {author} {\bibfnamefont {J.}~\bibnamefont
  {Blume}}, \bibinfo {author} {\bibfnamefont {Y.~H.}\ \bibnamefont {Jiang}},
  \bibinfo {author} {\bibfnamefont {P.}~\bibnamefont {Chen}},\ and\ \bibinfo
  {author} {\bibfnamefont {M.}~\bibnamefont {Weidem\"uller}},\ }\bibfield
  {title} {\bibinfo {title} {Two-dimensional magneto-optical trap as a source
  for cold strontium atoms},\ }\href
  {https://doi.org/10.1103/PhysRevA.96.053415} {\bibfield  {journal} {\bibinfo
  {journal} {Phys. Rev. A}\ }\textbf {\bibinfo {volume} {96}},\ \bibinfo
  {pages} {053415} (\bibinfo {year} {2017})}\BibitemShut {NoStop}%
\bibitem [{\citenamefont {Barbiero}\ \emph {et~al.}(2020)\citenamefont
  {Barbiero}, \citenamefont {Tarallo}, \citenamefont {Calonico}, \citenamefont
  {Levi}, \citenamefont {Lamporesi},\ and\ \citenamefont
  {Ferrari}}]{barbiero2020}%
  \BibitemOpen
  \bibfield  {author} {\bibinfo {author} {\bibfnamefont {M.}~\bibnamefont
  {Barbiero}}, \bibinfo {author} {\bibfnamefont {M.~G.}\ \bibnamefont
  {Tarallo}}, \bibinfo {author} {\bibfnamefont {D.}~\bibnamefont {Calonico}},
  \bibinfo {author} {\bibfnamefont {F.}~\bibnamefont {Levi}}, \bibinfo {author}
  {\bibfnamefont {G.}~\bibnamefont {Lamporesi}},\ and\ \bibinfo {author}
  {\bibfnamefont {G.}~\bibnamefont {Ferrari}},\ }\bibfield  {title} {\bibinfo
  {title} {Sideband-enhanced cold atomic source for optical clocks},\ }\href
  {https://doi.org/10.1103/PhysRevApplied.13.014013} {\bibfield  {journal}
  {\bibinfo  {journal} {Phys. Rev. Appl.}\ }\textbf {\bibinfo {volume} {13}},\
  \bibinfo {pages} {014013} (\bibinfo {year} {2020})}\BibitemShut {NoStop}%
\bibitem [{\citenamefont {Farolfi}\ \emph {et~al.}(2019)\citenamefont
  {Farolfi}, \citenamefont {Trypogeorgos}, \citenamefont {Colzi}, \citenamefont
  {Fava}, \citenamefont {Lamporesi},\ and\ \citenamefont
  {Ferrari}}]{Magnetic_shield}%
  \BibitemOpen
  \bibfield  {author} {\bibinfo {author} {\bibfnamefont {A.}~\bibnamefont
  {Farolfi}}, \bibinfo {author} {\bibfnamefont {D.}~\bibnamefont
  {Trypogeorgos}}, \bibinfo {author} {\bibfnamefont {G.}~\bibnamefont {Colzi}},
  \bibinfo {author} {\bibfnamefont {E.}~\bibnamefont {Fava}}, \bibinfo {author}
  {\bibfnamefont {G.}~\bibnamefont {Lamporesi}},\ and\ \bibinfo {author}
  {\bibfnamefont {G.}~\bibnamefont {Ferrari}},\ }\bibfield  {title} {\bibinfo
  {title} {Design and characterization of a compact magnetic shield for
  ultracold atomic gas experiments},\ }\href
  {https://doi.org/10.1063/1.5119915} {\bibfield  {journal} {\bibinfo
  {journal} {Rev. Sci. Instrum.}\ }\textbf {\bibinfo {volume} {90}},\ \bibinfo
  {pages} {115114} (\bibinfo {year} {2019})}\BibitemShut {NoStop}%
\bibitem [{\citenamefont {Dedman}\ \emph {et~al.}(2007)\citenamefont {Dedman},
  \citenamefont {Dall}, \citenamefont {Bryon},\ and\ \citenamefont
  {Truscott}}]{Dedman}%
  \BibitemOpen
  \bibfield  {author} {\bibinfo {author} {\bibfnamefont {C.~J.}\ \bibnamefont
  {Dedman}}, \bibinfo {author} {\bibfnamefont {R.~G.}\ \bibnamefont {Dall}},
  \bibinfo {author} {\bibfnamefont {L.~J.}\ \bibnamefont {Bryon}},\ and\
  \bibinfo {author} {\bibfnamefont {A.~G.}\ \bibnamefont {Truscott}},\
  }\bibfield  {title} {\bibinfo {title} {Active cancellation of stray magnetic
  fields in a bose-einstein condensation experiment},\ }\href
  {https://doi.org/10.1063/1.2472600} {\bibfield  {journal} {\bibinfo
  {journal} {Rev. Sci. Instrum.}\ }\textbf {\bibinfo {volume} {78}},\ \bibinfo
  {pages} {024703} (\bibinfo {year} {2007})}\BibitemShut {NoStop}%
\bibitem [{\citenamefont {Ringot}\ \emph {et~al.}(2001)\citenamefont {Ringot},
  \citenamefont {Szriftgiser},\ and\ \citenamefont {Garreau}}]{Ringot2001}%
  \BibitemOpen
  \bibfield  {author} {\bibinfo {author} {\bibfnamefont {J.}~\bibnamefont
  {Ringot}}, \bibinfo {author} {\bibfnamefont {P.}~\bibnamefont
  {Szriftgiser}},\ and\ \bibinfo {author} {\bibfnamefont {J.~C.}\ \bibnamefont
  {Garreau}},\ }\bibfield  {title} {\bibinfo {title} {Subrecoil raman
  spectroscopy of cold cesium atoms},\ }\href
  {https://doi.org/10.1103/PhysRevA.65.013403} {\bibfield  {journal} {\bibinfo
  {journal} {Phys. Rev. A}\ }\textbf {\bibinfo {volume} {65}},\ \bibinfo
  {pages} {013403} (\bibinfo {year} {2001})}\BibitemShut {NoStop}%
\bibitem [{\citenamefont {Botti}\ \emph {et~al.}(2006)\citenamefont {Botti},
  \citenamefont {Buffa}, \citenamefont {Bertoldi}, \citenamefont {Bassi},\ and\
  \citenamefont {Ricci}}]{botti2006}%
  \BibitemOpen
  \bibfield  {author} {\bibinfo {author} {\bibfnamefont {L.}~\bibnamefont
  {Botti}}, \bibinfo {author} {\bibfnamefont {R.}~\bibnamefont {Buffa}},
  \bibinfo {author} {\bibfnamefont {A.}~\bibnamefont {Bertoldi}}, \bibinfo
  {author} {\bibfnamefont {D.}~\bibnamefont {Bassi}},\ and\ \bibinfo {author}
  {\bibfnamefont {L.}~\bibnamefont {Ricci}},\ }\bibfield  {title} {\bibinfo
  {title} {Noninvasive system for the simultaneous stabilization and control of
  magnetic field strength and gradient},\ }\href
  {https://doi.org/10.1063/1.2173846} {\bibfield  {journal} {\bibinfo
  {journal} {Rev. Sci. Instrum.}\ }\textbf {\bibinfo {volume} {77}},\ \bibinfo
  {pages} {035103} (\bibinfo {year} {2006})},\ \Eprint
  {https://arxiv.org/abs/https://doi.org/10.1063/1.2173846}
  {https://doi.org/10.1063/1.2173846} \BibitemShut {NoStop}%
\bibitem [{\citenamefont {Smith}\ \emph {et~al.}(2011)\citenamefont {Smith},
  \citenamefont {Anderson}, \citenamefont {Chaudhury},\ and\ \citenamefont
  {Jessen}}]{Smith2011}%
  \BibitemOpen
  \bibfield  {author} {\bibinfo {author} {\bibfnamefont {A.}~\bibnamefont
  {Smith}}, \bibinfo {author} {\bibfnamefont {B.~E.}\ \bibnamefont {Anderson}},
  \bibinfo {author} {\bibfnamefont {S.}~\bibnamefont {Chaudhury}},\ and\
  \bibinfo {author} {\bibfnamefont {P.~S.}\ \bibnamefont {Jessen}},\ }\bibfield
   {title} {\bibinfo {title} {Three-axis measurement and cancellation of
  background magnetic fields to less than 50 $\upmu${G} in a cold atom
  experiment},\ }\href {https://doi.org/10.1088/0953-4075/44/20/205002}
  {\bibfield  {journal} {\bibinfo  {journal} {J. Phys. B: At. Mol. Opt. Phys.}\
  }\textbf {\bibinfo {volume} {44}},\ \bibinfo {pages} {205002} (\bibinfo
  {year} {2011})}\BibitemShut {NoStop}%
\bibitem [{\citenamefont {Xiao}\ \emph {et~al.}(2020)\citenamefont {Xiao},
  \citenamefont {Wang}, \citenamefont {Guo}, \citenamefont {Zhu}, \citenamefont
  {Zhao}, \citenamefont {Sun}, \citenamefont {Ye},\ and\ \citenamefont
  {Zhou}}]{Quieting_Bfield}%
  \BibitemOpen
  \bibfield  {author} {\bibinfo {author} {\bibfnamefont {K.}~\bibnamefont
  {Xiao}}, \bibinfo {author} {\bibfnamefont {L.}~\bibnamefont {Wang}}, \bibinfo
  {author} {\bibfnamefont {J.}~\bibnamefont {Guo}}, \bibinfo {author}
  {\bibfnamefont {M.}~\bibnamefont {Zhu}}, \bibinfo {author} {\bibfnamefont
  {X.}~\bibnamefont {Zhao}}, \bibinfo {author} {\bibfnamefont {X.}~\bibnamefont
  {Sun}}, \bibinfo {author} {\bibfnamefont {C.}~\bibnamefont {Ye}},\ and\
  \bibinfo {author} {\bibfnamefont {X.}~\bibnamefont {Zhou}},\ }\bibfield
  {title} {\bibinfo {title} {Quieting an environmental magnetic field without
  shielding},\ }\href {https://doi.org/10.1063/5.0007464} {\bibfield  {journal}
  {\bibinfo  {journal} {Rev. Sci. Instrum.}\ }\textbf {\bibinfo {volume}
  {91}},\ \bibinfo {pages} {085107} (\bibinfo {year} {2020})}\BibitemShut
  {NoStop}%
\bibitem [{\citenamefont {Pandey}\ \emph {et~al.}(2016)\citenamefont {Pandey},
  \citenamefont {Kwong}, \citenamefont {Pramod},\ and\ \citenamefont
  {Wilkowski}}]{pandey2016linear}%
  \BibitemOpen
  \bibfield  {author} {\bibinfo {author} {\bibfnamefont {K.}~\bibnamefont
  {Pandey}}, \bibinfo {author} {\bibfnamefont {C.~C.}\ \bibnamefont {Kwong}},
  \bibinfo {author} {\bibfnamefont {M.~S.}\ \bibnamefont {Pramod}},\ and\
  \bibinfo {author} {\bibfnamefont {D.}~\bibnamefont {Wilkowski}},\ }\bibfield
  {title} {\bibinfo {title} {Linear and nonlinear magneto-optical rotation on
  the narrow strontium intercombination line},\ }\href
  {https://doi.org/10.1103/PhysRevA.93.053428} {\bibfield  {journal} {\bibinfo
  {journal} {Phys. Rev. A}\ }\textbf {\bibinfo {volume} {93}},\ \bibinfo
  {pages} {053428} (\bibinfo {year} {2016})}\BibitemShut {NoStop}%
\bibitem [{\citenamefont {Kwong}(2017)}]{kwong2017}%
  \BibitemOpen
  \bibfield  {author} {\bibinfo {author} {\bibfnamefont {C.~C.}\ \bibnamefont
  {Kwong}},\ }\emph {\bibinfo {title} {Coherent transmission of light through a
  cold atomic cloud}},\ \href {https://doi.org/10.32657/10356/69897} {Ph.D.
  thesis},\ \bibinfo  {school} {Nanyang Technological University} (\bibinfo
  {year} {2017})\BibitemShut {NoStop}%
\bibitem [{\citenamefont {Kwiatkowski}\ and\ \citenamefont
  {Tumanski}(1986)}]{Kwiatkowski1986}%
  \BibitemOpen
  \bibfield  {author} {\bibinfo {author} {\bibfnamefont {W.}~\bibnamefont
  {Kwiatkowski}}\ and\ \bibinfo {author} {\bibfnamefont {S.}~\bibnamefont
  {Tumanski}},\ }\bibfield  {title} {\bibinfo {title} {The permalloy
  magnetoresistive sensors-properties and applications},\ }\href
  {https://doi.org/10.1088/0022-3735/19/7/002} {\bibfield  {journal} {\bibinfo
  {journal} {J. Phys. E: Sci. Instr.}\ }\textbf {\bibinfo {volume} {19}},\
  \bibinfo {pages} {502} (\bibinfo {year} {1986})}\BibitemShut {NoStop}%
\bibitem [{\citenamefont {Bertoldi}\ \emph {et~al.}(2005)\citenamefont
  {Bertoldi}, \citenamefont {Bassi}, \citenamefont {Ricci}, \citenamefont
  {Covi},\ and\ \citenamefont {Varas}}]{bertoldi2005}%
  \BibitemOpen
  \bibfield  {author} {\bibinfo {author} {\bibfnamefont {A.}~\bibnamefont
  {Bertoldi}}, \bibinfo {author} {\bibfnamefont {D.}~\bibnamefont {Bassi}},
  \bibinfo {author} {\bibfnamefont {L.}~\bibnamefont {Ricci}}, \bibinfo
  {author} {\bibfnamefont {D.}~\bibnamefont {Covi}},\ and\ \bibinfo {author}
  {\bibfnamefont {S.}~\bibnamefont {Varas}},\ }\bibfield  {title} {\bibinfo
  {title} {Magnetoresistive magnetometer with improved bandwidth and response
  characteristics},\ }\href {https://doi.org/10.1063/1.1922787} {\bibfield
  {journal} {\bibinfo  {journal} {Rev. Sci. Instrum.}\ }\textbf {\bibinfo
  {volume} {76}},\ \bibinfo {pages} {065106} (\bibinfo {year} {2005})},\
  \Eprint {https://arxiv.org/abs/https://doi.org/10.1063/1.1922787}
  {https://doi.org/10.1063/1.1922787} \BibitemShut {NoStop}%
\bibitem [{\citenamefont {{Bertoldi, A.}}\ \emph {et~al.}(2006)\citenamefont
  {{Bertoldi, A.}}, \citenamefont {{Botti, L.}}, \citenamefont {{Covi, D.}},
  \citenamefont {{Buffa, R.}}, \citenamefont {{Bassi, D.}},\ and\ \citenamefont
  {{Ricci, L.}}}]{bertoldi2006}%
  \BibitemOpen
  \bibfield  {author} {\bibinfo {author} {\bibnamefont {{Bertoldi, A.}}},
  \bibinfo {author} {\bibnamefont {{Botti, L.}}}, \bibinfo {author}
  {\bibnamefont {{Covi, D.}}}, \bibinfo {author} {\bibnamefont {{Buffa, R.}}},
  \bibinfo {author} {\bibnamefont {{Bassi, D.}}},\ and\ \bibinfo {author}
  {\bibnamefont {{Ricci, L.}}},\ }\bibfield  {title} {\bibinfo {title} {Noise
  and response characterization of an anisotropic magnetoresistive sensor
  working in a high-frequency flipping regime},\ }\href
  {https://doi.org/10.1051/epjap:2005088} {\bibfield  {journal} {\bibinfo
  {journal} {Eur. Phys. J. Appl. Phys.}\ }\textbf {\bibinfo {volume} {33}},\
  \bibinfo {pages} {51} (\bibinfo {year} {2006})}\BibitemShut {NoStop}%
\bibitem [{\citenamefont {Kwong}\ \emph {et~al.}(2014)\citenamefont {Kwong},
  \citenamefont {Yang}, \citenamefont {Pramod}, \citenamefont {Pandey},
  \citenamefont {Delande}, \citenamefont {Pierrat},\ and\ \citenamefont
  {Wilkowski}}]{PhysRevLett.113.223601}%
  \BibitemOpen
  \bibfield  {author} {\bibinfo {author} {\bibfnamefont {C.~C.}\ \bibnamefont
  {Kwong}}, \bibinfo {author} {\bibfnamefont {T.}~\bibnamefont {Yang}},
  \bibinfo {author} {\bibfnamefont {M.~S.}\ \bibnamefont {Pramod}}, \bibinfo
  {author} {\bibfnamefont {K.}~\bibnamefont {Pandey}}, \bibinfo {author}
  {\bibfnamefont {D.}~\bibnamefont {Delande}}, \bibinfo {author} {\bibfnamefont
  {R.}~\bibnamefont {Pierrat}},\ and\ \bibinfo {author} {\bibfnamefont
  {D.}~\bibnamefont {Wilkowski}},\ }\bibfield  {title} {\bibinfo {title}
  {Cooperative emission of a coherent superflash of light},\ }\href
  {https://doi.org/10.1103/PhysRevLett.113.223601} {\bibfield  {journal}
  {\bibinfo  {journal} {Phys. Rev. Lett.}\ }\textbf {\bibinfo {volume} {113}},\
  \bibinfo {pages} {223601} (\bibinfo {year} {2014})}\BibitemShut {NoStop}%
\bibitem [{\citenamefont {Hobson}\ \emph {et~al.}(2021)\citenamefont {Hobson},
  \citenamefont {Vovrosh}, \citenamefont {Stray}, \citenamefont {Packer},
  \citenamefont {Winch}, \citenamefont {Holmes}, \citenamefont {Hayati},
  \citenamefont {McGovern}, \citenamefont {Bowtell}, \citenamefont {Brookes}
  \emph {et~al.}}]{hobson2021bespoke}%
  \BibitemOpen
  \bibfield  {author} {\bibinfo {author} {\bibfnamefont {P.}~\bibnamefont
  {Hobson}}, \bibinfo {author} {\bibfnamefont {J.}~\bibnamefont {Vovrosh}},
  \bibinfo {author} {\bibfnamefont {B.}~\bibnamefont {Stray}}, \bibinfo
  {author} {\bibfnamefont {M.}~\bibnamefont {Packer}}, \bibinfo {author}
  {\bibfnamefont {J.}~\bibnamefont {Winch}}, \bibinfo {author} {\bibfnamefont
  {N.}~\bibnamefont {Holmes}}, \bibinfo {author} {\bibfnamefont
  {F.}~\bibnamefont {Hayati}}, \bibinfo {author} {\bibfnamefont
  {K.}~\bibnamefont {McGovern}}, \bibinfo {author} {\bibfnamefont
  {R.}~\bibnamefont {Bowtell}}, \bibinfo {author} {\bibfnamefont
  {M.}~\bibnamefont {Brookes}}, \emph {et~al.},\ }\bibfield  {title} {\bibinfo
  {title} {Bespoke magnetic field design for a magnetically shielded cold atom
  interferometer},\ }\href@noop {} {\bibfield  {journal} {\bibinfo  {journal}
  {arXiv preprint arXiv:2110.04498}\ } (\bibinfo {year} {2021})}\BibitemShut
  {NoStop}%
\bibitem [{\citenamefont {Hidalgo-Tobon}(2010)}]{hidalgo2010theory}%
  \BibitemOpen
  \bibfield  {author} {\bibinfo {author} {\bibfnamefont {S.~S.}\ \bibnamefont
  {Hidalgo-Tobon}},\ }\bibfield  {title} {\bibinfo {title} {Theory of gradient
  coil design methods for magnetic resonance imaging},\ }\href@noop {}
  {\bibfield  {journal} {\bibinfo  {journal} {Concepts in Magnetic Resonance
  Part A}\ }\textbf {\bibinfo {volume} {36}},\ \bibinfo {pages} {223} (\bibinfo
  {year} {2010})}\BibitemShut {NoStop}%
\bibitem [{\citenamefont {Hasan}\ \emph {et~al.}(2022)\citenamefont {Hasan},
  \citenamefont {Madasu}, \citenamefont {Rathod}, \citenamefont {Kwong},
  \citenamefont {Miniatura}, \citenamefont {Chevy},\ and\ \citenamefont
  {Wilkowski}}]{hasan2022anisotropic}%
  \BibitemOpen
  \bibfield  {author} {\bibinfo {author} {\bibfnamefont {M.}~\bibnamefont
  {Hasan}}, \bibinfo {author} {\bibfnamefont {C.~S.}\ \bibnamefont {Madasu}},
  \bibinfo {author} {\bibfnamefont {K.~D.}\ \bibnamefont {Rathod}}, \bibinfo
  {author} {\bibfnamefont {C.~C.}\ \bibnamefont {Kwong}}, \bibinfo {author}
  {\bibfnamefont {C.}~\bibnamefont {Miniatura}}, \bibinfo {author}
  {\bibfnamefont {F.}~\bibnamefont {Chevy}},\ and\ \bibinfo {author}
  {\bibfnamefont {D.}~\bibnamefont {Wilkowski}},\ }\bibfield  {title} {\bibinfo
  {title} {Wave packet dynamics in synthetic non-abelian gauge fields},\
  }\href@noop {} {\bibfield  {journal} {\bibinfo  {journal} {arXiv preprint
  arXiv:2201.00885}\ } (\bibinfo {year} {2022})}\BibitemShut {NoStop}%
\end{thebibliography}%
\end{document}